\documentclass[11pt,twoside]{article}
\usepackage[latin1]{inputenc}
\usepackage{graphicx}
\usepackage{rotating}
\usepackage{natbib}
\usepackage[english]{babel}
\usepackage{booktabs}
\usepackage{latexsym}
\usepackage{multirow}
\usepackage{authblk}
\usepackage[dvipsnames,usenames]{color}

\usepackage{amsmath, amsthm, amsfonts}
\usepackage{amssymb}

\date{}
\def \uu {\mathbf{u}}

\def \YY {\mathbf{Y}}

\def \XX {\mathbf{X}}

\def \VV {\mathbf{V}}

\def \ZZ {\mathbf{Z}}

\def \mmu {\boldsymbol{\mu}}

\def \bbeta {\boldsymbol{\beta}}

\def \SSigma {\boldsymbol{\Sigma}}

\graphicspath{ {./Figures/} }
\title{A new approach to the gender pay gap decomposition by economic activity\footnote{Supported by the MINECO grants MTM2017-82724-R, MTM2015-71217-R, and by the Xunta de Galicia (Grupos de Referencia Competitiva ED431C-2016-015 and Centro Singular de Investigaci\'on de Galicia ED431G/01), all of them through the ERDF.}}

\author{ Mar\'{\i}a Jos\'e Lombard\'{\i}a$^{1}$, Esther L\'opez-Vizca\'{\i}no$^{2}$\\
Cristina Rueda$^{3}$
\vspace{0.5cm}\\
{\small$^{1}$ Universidade da Coru\~na, CITIC, Spain,} \\
{\small$^{2}$ Instituto Galego de Estat\'{\i}stica, Spain,} \\
{\small $^{3}$Universidad de Valladolid, Spain}} {}

\begin{document}
\maketitle

{Abstract}\\

The aim of this paper is to present an original approach to estimate the gender pay gap. We propose a model-based decomposition, similar to the most popular approaches, where the first component measures differences in group characteristics and the second component measures the unexplained effect; the latter being the real gap. The novel approach incorporates model selection and bias correction.  %,  avoiding the main limitation of standard approaches, which is the dependence on the choice of explanatory variables and the functional form in regression.   
The pay gap problem in a small area context is considered in this paper, although  the approach is  flexible to be applied to other contexts.

Specifically, the methodology is validated for analysing wage differentials by economic activities in the region of Galicia (Spain) and by analysing simulated data from an experimental design that imitates the generation of real data. The good performance of the proposed estimators is shown in both cases, specifically when compared with those obtained from the widely used Oaxaca-Blinder approach. 

\section{Introduction}

Gender wage inequality is  a documented fact that happens in almost all of the industrialized countries (Juhn et al., 1993, Oaxaca and Ransom, 1994,1998, Al\'aez and Ullibarri, 2001, Moral-Arce et al., 2011, Fortin et al., 2011) and interest in the quantification of this gap has increased in recent years. 

In the case of the Spanish labor market, a number of actions were taken to provide women with real access to employment with full social and economic rights, with special emphasis on the reconciliation of family and working life. This and other reasons led a massive incorporation of women into the labor market, but it has not resulted in equality, that is,  women's unemployment rates are higher in most sectors, the jobs that women take do not involve the same degree of responsibility or decision-making power and women's participation is limited to a few sectors of the economy. But more important is that, for a similar job, it has been shown that men have better wages than women (Fern\'andez et al., 2000; Al\'aez and Ullibarri, 2001; Al\'aez et al., 2000-2003; De la Rica et al., 2008; Moral-Arce et al., 2011). 
\\

The gender pay gap ($GPG$), calculated as the difference between men's average hourly earnings and women's average hourly earnings, usually expressed as a percentage of the men's average hourly earnings,  is widely used as the key indicator to study the progress in the European Union (EU), see Leythienne and Ronkowski (2018), among others. Moreover, most researchers in this field (Jann, 2008, Moral-Arce et al., 2011, Anastasiade and Tillé, 2017a, 2017b, Hlavac, 2018) work with the transformed variable to logarithms, $Y=log(W)$, where $W$ is the hourly wages. 
The rationale behind the use of the logarithmic transformation is that the salaries have a skew distribution. However, the skewness is also exhibited on the log scale. Finding the optimal transformation is not straightforward. As an alternative, Graf et al. (2019) propose considering a much more flexible distribution called generalized beta of the second kind (GB2). 
In this work, we follow the ideas of most researchers. We are aware that it would be interesting to incorporate this proposal, but it goes beyond the objectives of this paper.\\

Typically the sample estimators are used directly as the GPG measure because the sample size are often very high. Instead, a more formal definition of the GPG is used here, as we deal with scenarios with small sample sizes. It is given, using expectations, as follows:

\begin{equation*}%\label{GPG}
GPG=\frac{E(W_m) - E(W_w)}{E(W_m)},
\end{equation*}

where  the subscripts $m$ are the men's wage and $w$ the women's wage.\\

The most widely used approach to analyze the gender gap is the decomposition due to Oaxaca (1973) and Blinder (1973); Specifically, Eurostat applies this approach to decompose the estimator of $GPG$ (Leythienne and Ronkowski, 2018). The Oaxaca-Blinder (OB) decomposition breaks down the difference $\Delta=E(Y_m)-E(Y_w)$  into two components: $\Delta=Q+U$, where the first component explains the difference between the observed productive characteristics, such as education or work experience, and the second accounts for differences in the structure of the model (Anastasiade and Tillé, 2017a, 2017b). The U component is known as the unexplained component and is usually considered to be the wage discrimination by gender in the labor market.  By instances, Popli (2013) states that the unexplained wage gap includes the effect of labour market discrimination, unobservable variables and omitted variables. The methodology presented in this paper accounts for the bias from unobservable or omitted variables. 

So, the $GPG$ decomposition is:

\begin{equation}\label{GPGdescomp}
{GPG}_{Q}={GPG} \frac{{Q}}{\Delta}\ \quad\ \mbox{and}\quad\
{GPG}_{U}={GPG} \frac{{U}}{\Delta},
\end{equation}

\noindent which represent the explained and unexplained part of the $GPG$, respectively. The OB method, including the definitions of  $Q$ and $U$, is briefly revised in Section 2.

Despite its widespread use, the standard Oaxaca-Blinder decomposition has important limitations; firstly, too simple parametric functions are often used for the regression model; secondly, the results may depend heavily on the set of explanatory variables selected; and thirdly, the unexplained variability is not always taken into account and  generates bias.
\\

This paper aims to adapt the methodology to estimate  $GPG_Q$ and $GPG_U$ in small areas, which are disaggregated levels of the population, and specifically analyze the wage differentials by economic activities in the region of Galicia (Spain). Often in small area contexts, the sample sizes are too small, which implies the standard estimators are unreliable. In particular, in our application for 25\% of the economic activities, we have information on less than 26 male workers, being the sample sizes for women even smaller. 

In fact, no official estimators of wages are provided at the economic activity level because direct estimates have very low accuracy. The latter is a typical SAE (Small Area Estimation) problem. The most extended SAE approach is to give a model-based estimator, where auxiliary information in terms of explanatory variables is incorporated. Besides, the model is often defined as a mixed effect model where a random effect represents the areas (see Rao and Molina(2015). 
\\

The proposal here is to combine SAE methodology and  OB methodology by deriving a decomposition inspired by (\ref{GPGdescomp}), adapted to a small area scenario in such a way that for a given area $d$, $GPG_d$ is also decomposing into the $GPG_{Qd}$ and $GPG_{Ud}$. These components are efficiently estimated using an approach in three steps that use SAE techniques and overcome OB's drawback, commented above. First, by doing model selection using an AIC criterion, and second, by including a bias correction term that accounts for omitted variable bias as in many applications, the implicit assumption that all confounding variables are included is not easy to attain. Confidence intervals for $GPG_{Qd}$ and $GPG_{Ud}$ are obtained for each area. 
\\

Furthermore, to validate the methodology proposal, a simulation study is conducted. It has a particular design that mimics the generation of data of the real application at hand. The estimators obtained with the novel approach are compared with those using the OB approach, showing that the formers outperform the latter. 
 \\

We organize the remainder of the paper as follows. Section 2 revises the OB approach and SAE models, while Section 3 details the novel proposal. The estimation of the gender gap by economic activities in the region of Galicia (Spain) is presented in Section 4 and the simulation study in Section 5. Finally, the main conclusions and a brief discussion of future research is given in Section 6.

\section{Background}

We use the notation which is usually considered for the OB decomposition and SAE models. The beta parameter appears in both formulations, but with a different role in each one.
\subsection{Pay gap decomposition}{\label{OB}}

Let $Y_{g}=log(W_{g}) $ be the natural log of hourly earnings for $g=m, w$; men and women, respectively. And let  $\XX_{g}$ ;$(X_{g1},\ldots, X_{gp})$ represent the set of available explanatory variables. Typically, these explanatory variables are the education level, the age of the worker or the work experience.
\\

Assume that samples of sizes $n_m, n_w$ are available for men and women, respectively, and the following models:
\begin{equation*}%\label{ecuacion2}
Y_{gi}=\XX_{gi}\bbeta_g+\epsilon_{gi},\quad i=1,...,n_g,\quad g\in(m,w);
\end{equation*}
\noindent where
 $\bbeta_g=(\beta_{g1},\ldots, \beta_{gp})'$ is an unknown vector of parameters and $\epsilon_{gi}$ is independent, with $E(\epsilon_{g})=0$. 
  
Then, 

\begin{equation*} %\label{ecuacion3}
\Delta = E({Y}_{m})-E({Y}_{w})=\overline{\XX}_{m}\bbeta_m - \overline{\XX}_{w}\bbeta_w,
\end{equation*}
where, $\bar \XX_m$ and $\bar \XX_w$ are the corresponding sample means. 

Now, let $\beta_r$ be the auxiliary coefficient; then,  adding and subtracting $\overline{\XX}_{m}\bbeta_r$ and $\overline{\XX}_{w}\bbeta_r$, the  above difference can be written as:

\begin{equation*}%\label{ecuacion5}
\Delta =(\overline{\XX}_{m}-\overline{\XX}_{w})\bbeta_r+\overline{\XX}_{m}(\bbeta_m-\bbeta_r) + \overline{\XX}_{w}(\bbeta_r-\bbeta_w).
\end{equation*}
Frequently, the discrimination is directed toward one of the groups (women, in our case), which determines the value of the auxiliary coefficient, so that $\beta_r=\beta_m$ and then,

\begin{equation*}%\label{ecuacion5}
\Delta =(\overline{\XX}_{m}-\overline{\XX}_{w})\bbeta_m+\overline{\XX}_{w}(\bbeta_m-\bbeta_w).
\end{equation*}
 
where
$$
Q=(\overline{\XX}_{m}-\overline{\XX}_{w})\bbeta_m,
$$
\noindent is the component that is explained by group differences in the predictors (the ``quantify effect'' or ``the explained part'') and  

$$
U=\overline{\XX}_{w}(\bbeta_m-\bbeta_w).
$$

\noindent is the unexplained component, usually attributed to discrimination. Notice  that $U$ also captures all the potential effects of differences in unobserved variables.
\\

The estimation of the decomposition is straightforward. Let $\hat \bbeta_m$ and $\hat \bbeta_w$ be the least-squares estimates for $\bbeta_m$ and $\bbeta_w$ obtained separately from the two samples.  Then, 
\begin{eqnarray*}%\label{hatDelta}
\begin{array}{lll}
\hat\Delta &=& {\bar{Y}}_{m}-{\bar{Y}}_{w}= \hat{Q}+\hat{U},\\
\hat{Q}&=&(\bar\XX_{m}-\bar\XX_{w})\hat\bbeta_m,\\
\hat{U}&=& \bar\XX_{w}^{'}(\hat\bbeta_m-\hat\bbeta_w).
\end{array}
\end{eqnarray*}
%where 
%$ \hat{Q}=(\bar\XX_{m}-\bar\XX_{w})\hat\bbeta_m={\bar{Y}}_{m}-{\bar{Y}}_{w}^{m}$, and $\hat{U}= \bar\XX_{w}^{'}(\hat\bbeta_m-\hat\bbeta_w)={\bar{Y}}_{w}^{m}-{\bar{Y}}_{w}$
%Notice that  ${\bar{Y}}_{w}^{m}=\bar\XX_{w}\hat\bbeta_m$ is the estimator of the $E(Y_w)$ using the parameter estimated from the men fitted model.

More details and other decompositions can be seen in Jann (2008) and Hlavac (2018).
\\
 \subsection{SAE models}

The wage estimation by small areas is a typical SAE problem as it is assumed that the sample sizes are too low in many areas. The most extended approach in SAE problems is to give a model-based estimator that usually includes random effects due to the areas. To clarify the presentation we have only considered linear models, as these are widely used in SAE applications and are well suited to analyzing the problem at hand. Nevertheless, the proposal in the paper can be adapted to consider other additive non-linear models  as those  which include monotone and spline models such as those in Opsomer et al. (2008) and Rueda and Lombard\'{\i}a (2012).  
\\

Specifically, the so-called nested error regression model with sampling weights, which, for the $i$-th worker in area $d$, is defined as follows:

\begin{equation}\label{ecuacion8}
 Y_{di} =\XX_{di}\bbeta_d+\ZZ_{di}\uu_{d}+\epsilon_{di}, d=1,\ldots, D, i=1,\ldots, n_d;
\end{equation}

\noindent where $Y_{di}$ is the (log) wage,  $\XX_{di}=(X_{1di},\ldots, X_{pdi})$ is a $p$ vector of explanatory variables, $\bbeta_d$ is the vector of unknown parameters; $\ZZ_{di}\uu_{d}$ are the random effects, with $\ZZ_{di}$ being a $q$ dimensional vector also defined from the explanatory variables, and $\uu_{d}\in N(0,\SSigma_u)$, where $\SSigma_u$ is a matrix of dimension $q$. $\uu_d$ is  independent of the model error $\epsilon_{di}$, which are assumed to be independent $N(0,\sigma_e^2 w_{di}^{-1})$, where $w_{di}$ is the sampling weight for individual $i$ in area $d$. In these models the covariance of the response variable in area $d$ is $\VV_{d}=\ZZ_d \SSigma_u \ZZ_d+\SSigma_e$ where $ \SSigma_e=\sigma_e^2{W_d}^{-1}$ and ${W_d}=diag(w_{di})_{n_d \times n_d}$. 
%The covariance of $\YY$ is $\VV_Y=\underset{1\leq d \leq D}{\hbox{diag}}{(\VV_d)}$. 
Two particular interesting cases are considered by Battese et al. (1988) and Dempster et al. (1981).
\\

In SAE applications, researchers are mainly interested in deriving estimators for the response variable in small areas; they usually work with the conditional means as the parameters of interest, which are defined as follows:

\begin{equation*}%\label{mu_d}
\mu_d=E(Y_{d}|\uu_{d})=\bar\XX_{d}\bbeta_{d}+\bar\ZZ_d\uu_{d},d=1,\ldots, D
\end{equation*}

%If we consider the marginal approach

%\begin{equation}\label{tita_d}
%\ttheta_d=E(Y_{d})=\bar\XX_{d}\bbeta_{d},d=1,\ldots, D
%\end{equation}

%with $\VV_{d}=Var(\YY_d)=\ZZ_d \SSigma_u \ZZ_d+\Sigma_e$ and $ \SSigma_e=\sigma_e^2I_{n_d}$
%\\

The estimation process starts with the variance components $(\SSigma_{u}, \SSigma_{e})$, which are estimated  by the residual maximum likelihood (REML) method;  the empirical best linear unbiased estimators  of $\bbeta_{d}$ and the empirical best linear unbiased predictors of $\uu_{d}$ are then obtained.
Furthermore, estimators for $\mu_d$ are derived as follows:

\begin{equation*}%\label{hatmu_d}
\hat{\mu}_d=\bar\XX_{d}\hat{\bbeta}_{d}+\bar\ZZ_d \hat{\uu}_{d},d=1,\ldots, D
\end{equation*}
 For details on the estimation process and further learning about SAE methods, the reader is referred to the  monographs of Jiang (2007) and  Rao and Molina (2015).\\

\section{The novel proposal}{\label{Models}}

Assume that we have wage data for men and women and explanatory variables from samples from $D$ small areas, and that we are interested in deriving estimates for the two components of the $GPG$.
%There are different ways of adapting the Oaxaca-Blinder decomposition to derive pay gap estimators in a small area context.

A three-step approach is proposed to derive GPG estimators. The first step is a model selection step; in the second step, preliminary area-specific estimators for $GPG_Q$ and $GPG_U$ are obtained from a GPG decomposition adapted to small areas; and finally, in the third step, a Monte-carlo algorithm is consider to derive corrected bias $GPG_Q$ and $GPG_U$ estimators and confidence intervals.
%In our proposal, the estimators are obtained from a two step approach. In the first step,  a set of models is proposed from which one is chosen using an AIC statistic. In a second step, area-specific estimators for $GPG_Q$ and $GPG_U$ are obtained from the selected model, using a combination of SAE methods and a Montecarlo bias correction approach. We detail both steps below.
\\

 \textbf{
First Step: Model Selection
}

There are different approaches to model selection, and the task is complicated when random effects are present. The discussions by Lombardia et al. (2017), where the GAIC is first presented, and by Fan and Li (2012), where iterative regularization methods are proposed, are very instructive.
Besides proposing a method for making the model selection, a list of candidate models must be defined. From the methodological point of view, it can be as large as the practitioner decides. From the computational and interpretative points of view, a reduced list of candidate models are recommended.

The strategy adopted in this paper is to define a reduced list of candidate models and then to use the GAIC statistic to select a model from this list. In particular, for applications with a reduced number and well-known, explanatory variables, we recommend the use of \textit{prior knowledge}  to decide an initial selection of fixed and random effects.  Typically, in small area applications, the variable determining the areas is modelled using random effects. Other candidate models may be defined by considering the interactions of that variable, with other explanatory variables.
\\

The group (men/women) with the larger sample size is used to select the model. In $GPG$ applications, it is usually that of men ($m$). The models from which we make the choice differ in the explanatory variables selected and the type of effects (fixed or random) that describe the model and are formulated as the general model (\ref{ecuacion8}), as follows:
\begin{equation*}%\label{ec_mw}
 M: Y_{mdi} =\XX_{mdi}\bbeta_{md}+\ZZ_{mdi}\uu_{md}+\epsilon_{mdi}, \quad\ d=1,\ldots, D, i=1,\ldots, n_d.
\end{equation*}

The selection of the best model is done using an $AIC$ statistic (Akaike, 1973); specifically, we use an adapted version for unit level models of the $xGAIC$ proposed by Lombard\'{\i}a et al. (2017) with the following expression for the model $M$

\begin{equation*}%\label{aic}
 xGAIC =-2log(l_x(M))+xGDF
\end{equation*}

\noindent where $l_x(M)$ is the quasi-loglikelihood of the model $M$, which considers the focus on the random effect and the total variability and $xGDF$ is a measure of the complexity of the model, as follows:

$$
\log (l_{x}(M))=-\frac{1}{2}D\log(2\pi)-\frac{1}{2}\log|\VV_{Y_m}|-\frac{1}{2}(\YY_m-\mmu_m)^{'}\VV_{Y_m}^{-1}(\YY_m-\mmu_m)
$$

and
\noindent 

$$
xGDF=\sum_{d=1}^D \frac{\partial E(\widehat \mmu_{md})}{\partial\mmu_{md}}=\sum_{d=1}^D \sum_{i=1}^{n_{md}}  \sum_{j=1}^{n_{md}} V_{mdij}^{-1} Cov(\widehat \mu_{mdi},Y_{mdj})
$$

\noindent where $\YY_m$ is the $\sum_{d=1}^D n_d × 1$ vector with elements $Y_{mdi}$,
$\mmu_m$ is the vector of the conditional means $\mmu_m=E(\YY_m|\uu)$ with  elements $\mmu_{mdi}=\XX_{mdi}\bbeta_{md}+\ZZ_{mdi}\uu_{md}$, 
%$\VV_{Y_m}=$\underset{1\leq d \leq D}{\hbox{diag}}{(\VV_{md})}$ the conditional covariance of $\YY_m$, following the expression $V_d$ in Section 2.2.
$\VV_{Y_m}$ is the conditional covariance matrix, which is diagonal with elements $\VV_{md}$ according to Section 2.2 and
$V_{mdij}^{-1}$  is the $ij$-element of $\VV_{md}^{-1}$.

 Parametric bootstrap is used to estimate  $xGDF$,  as the analytic values are difficult to obtain in a similar way to 
Lombard\'{\i}a et al. (2017). The model with the lowest value of $xGAIC$ is selected.
 \\
  
  \textbf{
Second Step: GPG decomposition. Preliminary estimators.}

 Let $X_m, Z_m$ and $X_w$,$Z_w$ be the design matrices representing the explanatory variables and the random effects on the selected model for men and women, respectively. For a given area, $d$, consider   $\Delta_d=\mu_{md}-\mu_{wd}$, which is decomposed as follows: 
 
 \begin{equation*} %\label{Delta_d}
\Delta_d=\mu_{md}-\mu_{wd}=\bar\XX_{md}\bbeta_{md}+\bar\ZZ_{md}\uu_{md}-\bar\XX_{wd}\bbeta_{wd}-\bar\ZZ_{wd}\uu_{wd}. 
\end{equation*}

\begin{eqnarray*}%\label{DeltaEst}
\begin{array}{lll}
\Delta_d & = & Q_d+U_d\\
{Q}_d&=&(\bar\XX_{md}-\bar\XX_{wd})\bbeta_{md}+(\bar\ZZ_{md}-\bar\ZZ_{wd})\uu_{md},\\
{U}_d&=& \bar\XX_{wd}(\bbeta_{md}-\bbeta_{wd})+\bar\ZZ_{wd}(\uu_{md}-\uu_{wd}).
\end{array}
\end{eqnarray*}
and,
\begin{equation*}%\label{GPGdescomp_d}
{GPG}_{Qd}={GPG}_{d} \frac{Q_d}{\Delta_{d}}\ \quad\ \mbox{and}\quad\
{GPG}_{Ud}={GPG}_{d} \frac{{U}_d}{\Delta_{d}}.
\end{equation*}
where

\begin{equation*}%\label{GPG_d}
GPG_d=\frac{E(W_{md}) - E(W_{wd})}{E(W_{md})},
\end{equation*}

Small area estimators for $Y_{gd}$, $\Delta_d$, $U_d$ and $Q_d$  (as in Rao and Molina, 2015) and estimators for  $E(W_{gd}), g \in (m,w)$ $d=1,...,D$, are defined as follows: 

\begin{eqnarray}\label{OBestimatiors_d}
\begin{array}{lll}
\hat\Delta_d & =& \bar\XX_{md}\hat\bbeta_{md}+\bar\ZZ_{md}\hat\uu_{md}-\bar\XX_{wd}\hat\bbeta_{wd}-\bar\ZZ_{wd}\hat\uu_{wd},\\
\hat{Q}_d&=&(\bar\XX_{md}-\bar\XX_{wd})\hat\bbeta_{md}+(\bar\ZZ_{md}-\bar\ZZ_{wd})\hat\uu_{md},\\
\hat{U}_d&=& \hat\Delta_d-\hat{Q}_d.
\end{array}
\end{eqnarray}

\begin{equation}\label{SAEestimador}
\widehat{E(W_{gd})} =\overline{ exp(\hat{Y}_{gd})}, g \in (m,w), d=1,...,D;
\end{equation}

 From (\ref{SAEestimador}), we have that:

\begin{equation}\label{hatGPG_d}
\widehat {GPG_d}=\frac{\widehat {E(W_{md})} - \widehat{E(W_{wd})}}{\widehat {E(W_{md})}},
\end{equation}

\textbf{
Third Step: bias corrected estimators and confidence intervals.}

 Omitted variable bias is taken into account as in many applications, the implicit assumption that all confounding variables are included is not easy to attain.  The bias in $Q_d$ is derived by splitting the men sample into two parts I times.  The value of $B_d^i$, for iteration $i=1,...,I$, is expected to be zero when comparing the populations representing these two parts. A non-zero estimated value is assumed to be due to the omitted variables and is denoted by $\hat{B}^i_d$, for iteration $i$. 
From these values, the bias in $Q_d$ is estimated, and bias-corrected estimators for  $GPG_{Qd}$  and $GPG_{Ud}$  are derived,  as explained below. Iterations with no men or women in any of the activities are discarded.  \\

For each iteration $i= 1, \ldots, I$ repeat: 
 
\begin{enumerate}

\item Randomly split the data $S$ in two parts of size $\frac{n}{2}$: $S_1^i$ and $S_2^{i}$. The individuals of $S_1^i$ are selected randomly, in such a way that those that have not been selected form part of $S_2^{i}$. Therefore, the union of these two sets is the complete sample. In the following, subindex 1 or 2 indicates the data coming from $S_1^i$ or $S_2^i$, respectively.

\item The model selected is fitted to men and women's data in  $S_1^i$, and $\hat{Q}_{1d}^i$,  $\hat{\Delta}_{1d}^i$, and  $\widehat{GPG}_{1d}^i$ are calculated according to (\ref{OBestimatiors_d}) and (\ref{hatGPG_d}).

\item For each area $d$ and iteration $i$, the bias is calculated considering men's data from $S_1^i$ and $S_2^i$, as follows:
$$
\hat{B}_{d}^{i}= (\bar\XX_{1md}-\bar\XX_{2md})\hat\bbeta_{1md}+(\bar\ZZ_{1md}-\bar\ZZ_{2md}) \hat u_{1md}.
$$

\end{enumerate}
Then, the final bias term is estimated as: 

\begin{equation*}%\label{bias_d}
\hat{B}_{d}=\frac{1}{I}\sum_{i=1}^{I}\hat{B}_{d}^{i},
\end{equation*}

\noindent and the corrected bias estimates of ${Q}_d$ and ${U}_d$ are:
$$
\hat{Q}_d^B= \hat{Q}_d+ \hat{B}_{d},
$$
$$
\hat{U}_d^B=\hat \Delta_d-\hat{Q}_d^B,
$$
where $\hat{Q}_d$ and $\hat{\Delta}_{d}$ are  from the complete sample, as in (\ref{OBestimatiors_d}).\\

Now, the corrected bias estimates of the explained and unexplained parts of $GPG$ are derived from the latter and from (\ref{hatGPG_d}) as:
\begin{equation}\label{GPGestimado_d}
\widehat{GPG}_{Qd}=\widehat{GPG}_{d} \frac{\hat{Q}_d^B}{{\hat\Delta_{d}}}\ \quad\ \mbox{and}\quad\
\widehat{GPG}_{Ud}=\widehat{GPG}_{d} \frac{\hat{U}_d^B}{{\hat\Delta_{d}}}
\end{equation}

Moreover, confidence intervals are also derived as follows:

%Then, the error of  $\widehat{GPG}_{Qd}^{B}$ can be calculed as

  From $S_1^i$, $i= 1, \ldots, I$, and (\ref{SAEestimador}), we have that:

$$\hat{Q}_d^{Bi}= \hat{Q}_{1d}^i+\hat{B}_{d},\quad\mbox{and} \quad \hat{U}_d^{Bi}= \hat \Delta_{1d}^{i}-\hat{Q}_d^{Bi}.$$

$$\widehat{GPG}_{1d}^i=\frac{\widehat {E(W_{1md})} - \widehat{E(W_{1wd})}}{\widehat {E(W_{1md})}},$$
$$\widehat{GPG}_{Qd}^{i}=\widehat{GPG}_{1d}^i \frac{\hat{Q}_d^{Bi}}{\hat\Delta_{1d}^i},$$
$$\widehat{GPG}_{Ud}^{i}=\widehat{GPG}_{1d}^i \frac{\hat{U}_d^{Bi}}{\hat\Delta_{1d}^i},$$

%\widehat{GPG}_{ud}^{Bi}=\widehat{GPG}_{di} \frac{\hat{U}_d^{Bi}}{\hat\Delta_d^i}.
%where $\widehat{GPG}_{d}^i$ and $\hat\Delta_{d}^i$ are the sampling estimators calculated with total data frame. 

Now, let us define:
\begin{eqnarray*}%{\label{MSE}}
\begin{array}{lll}
V(\widehat{GPG}_{Qd})&=&\frac{1}{I} \sum_{i=1}^{I}\left(\widehat{GPG}_{Qd}^{i}-\widehat{\overline{GPG}}_{Qd}^{B}\right)^2,\\
V(\widehat{GPG}_{Ud})&=&\frac{1}{I} \sum_{i=1}^{I}\left(\widehat{GPG}_{Ud}^{i}-\widehat{\overline{GPG}}_{Ud}^{B}\right)^2,
\end{array}
\end{eqnarray*}
and, $$\widehat{\overline{GPG}}_{Qd}^{B}=\frac{1}{I} \sum_{i=1}^{I}\widehat{GPG}_{Qd}^{i}.$$
 $$\widehat{\overline{GPG}}_{Ud}^{B}=\frac{1}{I} \sum_{i=1}^{I}\widehat{GPG}_{Ud}^{i}.$$ 
 
while the confidence intervals are derived as follows:

\begin{eqnarray}{\label{ICQU}}
\begin{array}{lll}
\widehat{{GPG}}_{Qd}\pm Z_{\alpha/2}\sqrt{V(\widehat{GPG}_{Qd})}\\
 \widehat{{GPG}}_{Ud}\pm Z_{\alpha/2}\sqrt{V(\widehat{GPG}_{Ud})}
\end{array}
\end{eqnarray}

The aim of this document is to provide an easy-to-apply methodology for constructing confidence intervals to allow for inferences to be made. The post-selection inference is an interesting issue that could be considered in the proposed methodology following the ideas of
Berk et al. (2013), Lee et al. (2016) and Tibshirani et al. (2016), among others; but it is outside the scope of this document.

\section{Estimation of the GPG by ecomomic activities in Galicia.}{\label{RealProblem}}

 The data come from the Structure of Earnings Survey (SES) conducted by the Spanish National Institute of Statistics (INE) in 2014. The subset considering in this paper, corresponds to the Galicia region at the Northwest of Spain. The survey is conducted in a similar way to wage structure surveys in other European countries; it uses a two-level sampling of national companies, a stratified sampling in the first stage for local units and systematic sampling to select workers at those units.
The economic activities are defined by the Statistical Classification of Economic Activities in the European Community (NACE09 Rev.2) at the division level.

Galicia has an economy strongly linked to natural resources, and its labor market has experienced growth in recent years. The $GPG$, it has been reduced from $15.5\%$ in 2010 to  $13.5\%$ in 2016. This latter value is below that of the EU ($16.3\%$), and below that of Spain ($15.1\%$) in that year.

The survey collects information on non-self-employed workers who work in establishments with at least 10 employees and covers a wide range of private sectors (industry, construction, commerce, catering business, transport, financial intermediation, \ldots) excluding the primary sector. %The information registered include education level, experience, type of contract ; a complete
The total sample size in Galicia being 10276.
The list of explanatory variables considered in this study are included in Table \ref{variables}, the selection being similar to that in Aláez et al., (2000, 2001, 2003), De la Rica et al. (2008) or  Moral-Arce et al. (2011).
% where $X_1$ is the {time worked in the current company} and $X_2$ the {age}. The qualitative auxiliary variables are:  {economic activity}, {size of the enterprise, market, occupation, type of work}, and {contract}. We define a design matrix $\XX$ where the columns $X_i, i=3,...104$ are binary variables for each of the k-1 last levels of the variables, taking the first category as the reference category. \\

The economic activities, which are defined by the Statistical Classification of Economic Activities in the European Community (NACE09 Rev.2) at the division level, are the areas of interest. Tables \ref{nace091} and \ref{nace092} in the Appendix include the list of codes and the description of the NACE09 at the division level, excluding the primary sector. In the present study, we have information for $D=78$ of the 84 NACE09's. SAE methods are appealing because the sample sizes in some activities are very small. Specifically, the  quartiles of the sample sizes  distribution in the areas are  $Q_{1}=26$, $Q_{2}=50$ and $Q_{3}=104$, for men, and  $Q_{1}=15$, $Q_{2}=26$ and $Q_{3}=50$, for women.\\ 
%being the  activity $7$, the reference in the analysis as no data for activity $6$ is recorder.  \\

Table \ref{characteristics} gives the main statistics for men and women regarding labor characteristics. The numbers in the table  show  a greater percentage  of women in higher education levels, 41\% against 33\% of men, and that almost half the men (48\%) are employed in the manufacturing industry or in public administration, while public administration (41\%) and the retail trade (16\%) are the activities with higher percentages for women.
Regarding their type of work, women are more frequently on part-time work (27\% vs 10\% for men). 

\begin{table}[h!]
\centering
\begin{tabular}{|lll|}
\hline
Name & Variable &	Description  	\\
\hline
\hline
Wage per hour& Y&	Gross hourly earnings from employment	\\
\hline
Gender &	&	Men, Women	\\
\hline
Experience& $X_1$&Number of years in the actual enterprise \\
\hline
Age&$X_2$	&Years of the employee	\\
\hline
Education	&&	(Primary) \\
 	&$X_{3}$&	 Secondary\\
    &$X_{4}$&	 Higher\\
\hline
Occupation	&&		(Professionals and managers)\\ 
&$X_{5}$&Tecnic \\ 
&$X_{6}$&Operators  \\ 
&$X_{7}$& Services and sales workers  \\ 
&$X_{8}$&Non skilled workers\\ 
\hline
Contract	&$X_{9}$&	 (Long term),  Short term\\
\hline
Type of work&$X_{10}$&	(Full time),  Part time\\
\hline
Size of the enterprise 	&&(Between 10 to 19 employees)  \\
&$X_{11}$&Between 20 to 49 employees \\ 
&$X_{12}$& Between  50 to 99 employees\\ 
&$X_{13}$& Between 100 to 199 employees  \\ 
&$X_{14}$&Between 200 to 499 employees  \\ 
&$X_{15}$& More than 500 workers  \\ 
\hline
Market	&&(Local or regional) \\
	&$X_{16}$&National \\
	&$X_{17}$&  UE \\
	&$X_{18}$& International\\
\hline
Aggregated economic activity&&(Energy)\\
& $X_{19}$ &Manufacturing industry \\ 
&$X_{20}$& Construction \\
&$X_{21}$& Retail trade  \\ 
&$X_{22}$&Transportation, storage and Accommodation \\ 
&$X_{23}$& Information and Communication\\
&$X_{24}$& Finance and insurance\\ 
&$X_{25}$& Professionals  \\ 
&$X_{26}$&  Public Administration \\
&$X_{27}$& Other service \\
\hline
Economic activity for two digits&{$X_{28}-X_{104}$}& {(7) NACE09 Rev. 2, division level, two digits}\\

\hline
\end{tabular}
\caption{\label{variables} Variables in the survey, in brackets the category used as a reference.
 For the variable  {\it Economic activity for two digits}, the reference category is the activity
7, because in Galicia there are not people working in the activity 6 (see Appendix). } 
\end{table}

\begin{table}[h!]
\centering
\begin{tabular}{|lrr|}
  \hline
 & Men & Women \\ 
  \hline
   \hline
Experience (mean years) &&\\
 \hline
  & 11.16& 10.43 \\ 
  \hline
  Education (\%)&&\\
  \hline
Primary & 0.19 & 0.13 \\ 
Secondary & 0.48 & 0.46 \\ 
Higher & 0.33 & 0.41 \\ 
  \hline
  Contract (\%) &&\\
  \hline
Long-term& 0.77 & 0.77 \\ 
Short-term& 0.23 & 0.23 \\ 
  \hline
Type of work (\%)&&\\
  \hline
Full time& 0.90 & 0.73 \\ 
Part time& 0.10 & 0.27 \\ 
  \hline
Economic activity (\%)&&\\
  \hline
Energy& 0.03 & 0.01 \\ 
Manufacturing industry& 0.27 & 0.13 \\ 
Construction& 0.09 & 0.01 \\ 
Retail trade & 0.13 & 0.16 \\ 
Transportation and storage and Accommodation & 0.08 & 0.07 \\ 
Information and Communication& 0.03 & 0.02 \\ 
Finance and insurance & 0.03 & 0.03 \\ 
Professionals & 0.10 & 0.14 \\ 
Public Administration & 0.21 & 0.41 \\ 
Other services & 0.02 & 0.02 \\ 
   \hline
\end{tabular}
\caption{\label{characteristics} Summary of labor force characteristics of the people in the survey by gender.}
\end{table}

\begin{figure}[!ht]
\begin{center}
\includegraphics[width=90mm,height=70 mm]{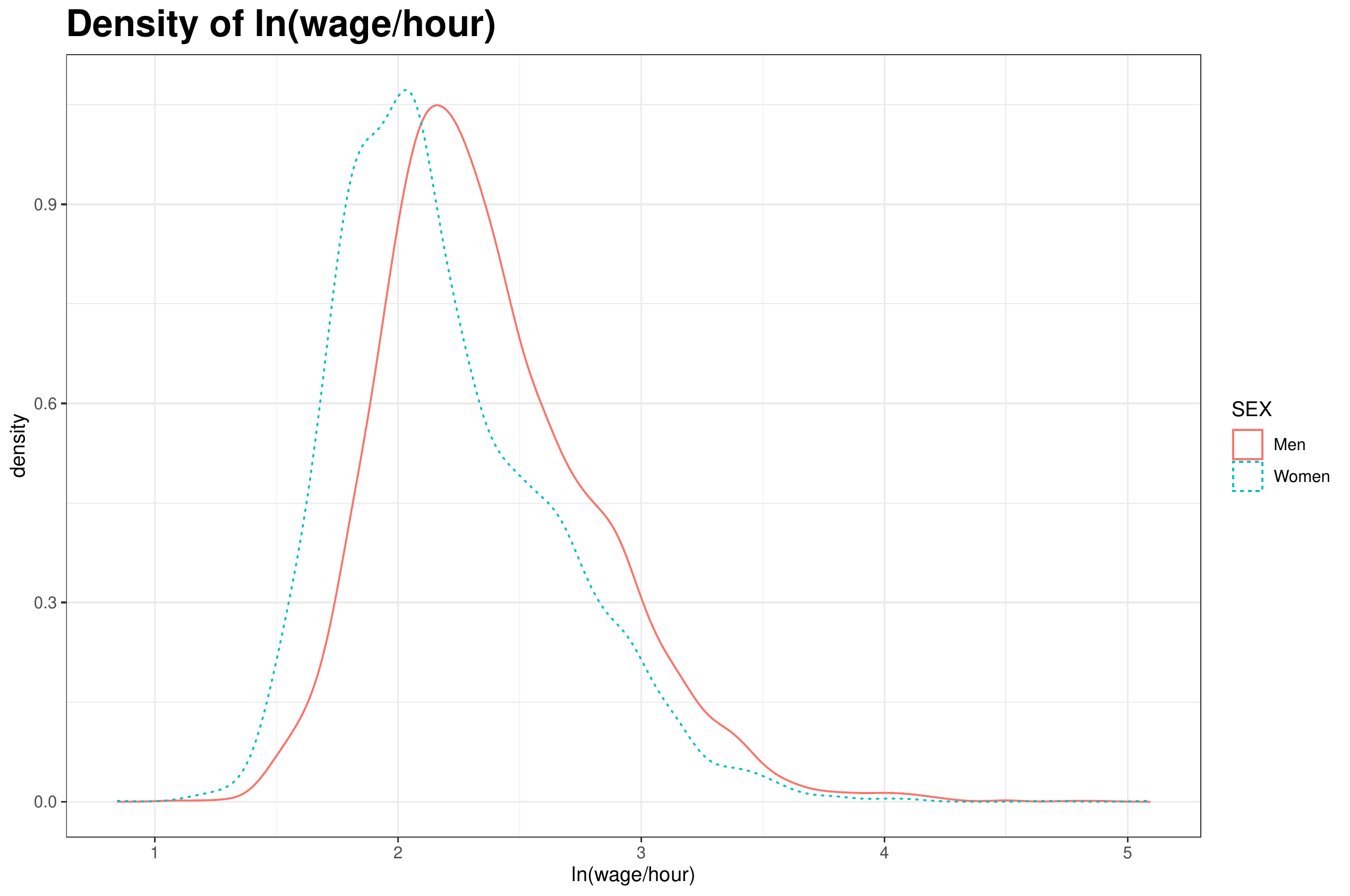}
\caption {\label{density} Density of ln(wage/hour) by sex in Galicia.}
\end{center}
\end{figure}

\begin{figure}[!ht]
\begin{center}
\includegraphics[width=160mm,height=90 mm]{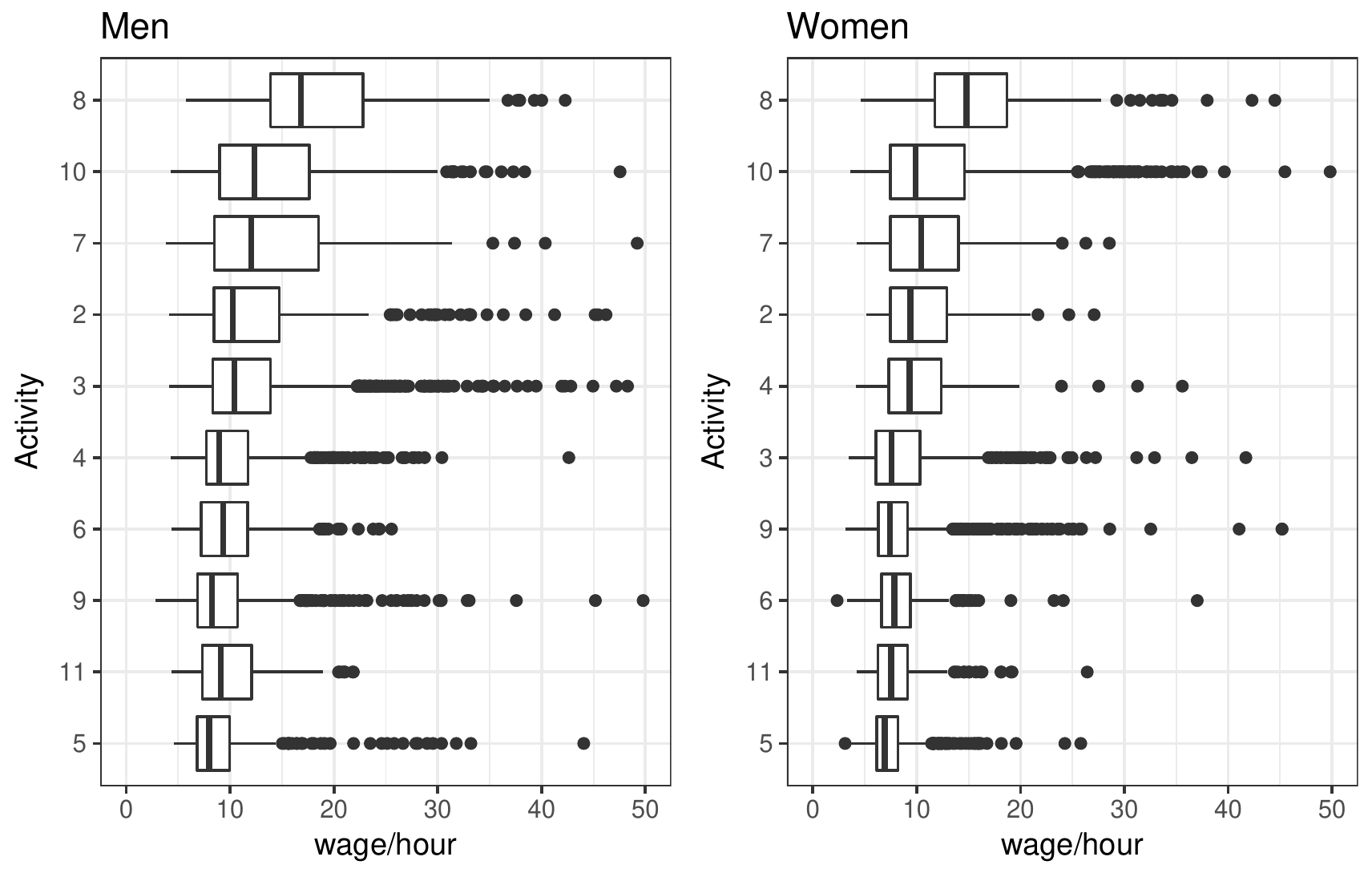}
\caption {\label{activity} Wage per hour by sex and economic activity in Galicia. The codes for the activities are: 2 Energy, 3 Manufacturing industry, 4 Construction, 5 Retail trade, 
6 Transportation and storage and Accommodation,
7 Information and Communication,
 8 Finance and insurance,
9 Professionals,
10 Public Administration and
 11 Other services. }
\end{center}
\end{figure}

The distribution of the response variable ($Y$), the wage per hour in logarithm, is shown in Figure \ref{density}  for men and women; the shape of both curves being quite similar, though that of men is shifted to the right, which indicates an increase in salaries. Specifically, the quartiles are Q1 = 7.6, Q2 = 9.6 and Q3 = 13.6 for men and Q1 = 6.4, Q2 = 8.1 and Q3 = 12.1 for women.
Finally, Figure \ref{activity} shows important differences in mean wages per hour across activities and between sexes. These differences range from 9 euros in the median of the activity \textit{retail trade} and the maximum of 18 euros for those employed in \textit{finance} and \textit{insurance} activities.
\\

The candidate list of models have been built as follows; First, the explanatory variables more often used in similar studies have been selected and have been described in Table \ref{variables}. Next, the candidate models are defined combining fixed effect for Experience, Age, Education, Occupation, Contract, Type of work, Size of the enterprise, and Market with the effects by the Economic Activities. The latter can be incorporated in the model using fixed effects for aggregated levels ($X19$ to $X27$), or fixed effects for each activity ($X28$ to $X104$), or  a random effect. Moreover, when a random effect is used to model the activities, the interactions with Experience, Education, and Occupation may also be considered.

Regarding the notation, $u_d$ is the random effect for the Economic Activities, $u_{dj},j=1, 2$ for  the interaction  with Education,  $u_{dk}$ $k=1, 2, 3, 4$ for the interaction with Occupation; and $v_d$ for the interaction with Experience. 
Only a selection of the most interesting models is presented in Table \ref{models} to simplify the exposition. Specifically, eight models labeled as $M1$ to $M8$ are presented, models $M6$ and $M7$ are defined with fixed effects, and the rest with a combination of fixed and random effects. Models defined by other combinations of effects have been considered but have been discarded because the results are far from being better than those obtained with other models.

{\tiny \begin{table}[h!]
\centering
\begin{tabular}{|llccc|}
\hline
 Label   &		Fixed effects	 &		Random Effects& Fixed parameters (nº) &Random parameters (nº)	\\
\hline
$M1$&$X_1$-$X_{18}$&$u_d$ &18&1\\
$M2$&$X_1$-$X_{18}$&$v_d$&18&1\\
$M3$&$X_1$-$X_{18}$&$u_d$, $v_d$&18&2\\
$M4$&$X_1$-$X_{18}$&$u_{dj}$&18&2\\
$M5$&$X_1$-$X_{18}$&$u_{dk}$&18&4\\
$M6$&$X_1$-$X_{27}$&-	&27&0\\
$M7$&$X_1$-$X_{18}$, $X_{28}$-$X_{104}$ &-&95&0\\
$M8$&$X_1$-$X_{18}$, $X_{28}$-$X_{104}$&$v_d$&95&1\\
   \hline
\end{tabular}
\caption{\label{models} Models used to fit the data.}
\end{table}

}

\begin{figure}[!ht]
\begin{center}
\includegraphics[width=170mm,height=90 mm]{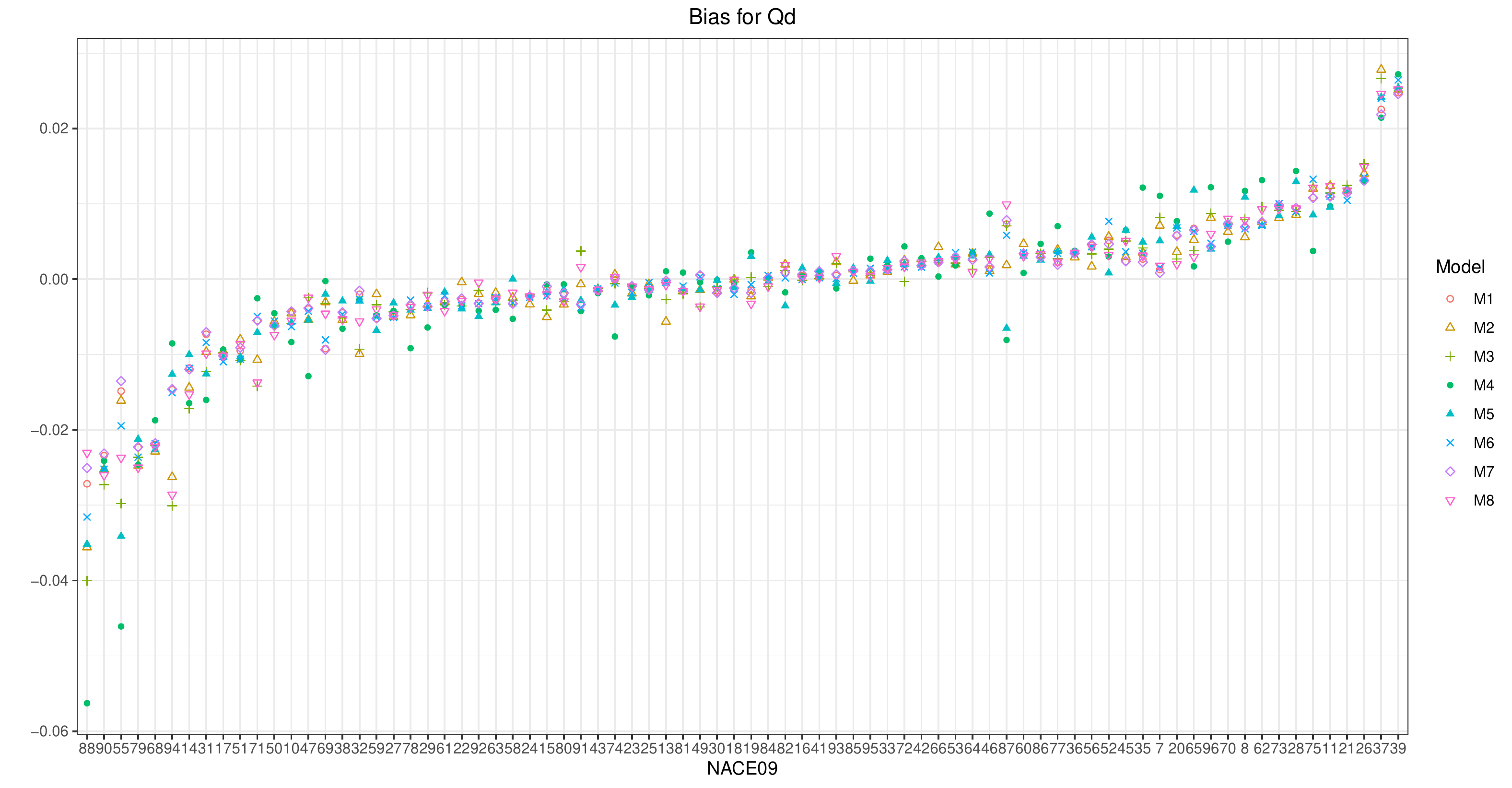}
\caption {\label{bias}Bias for $Q_d$ ($\hat B_d$), for each activity $d$ defined by the NACE09.}
\end{center}
\end{figure}

\begin{figure}[!ht]
\begin{center}
\includegraphics[width=180mm,height=90 mm]{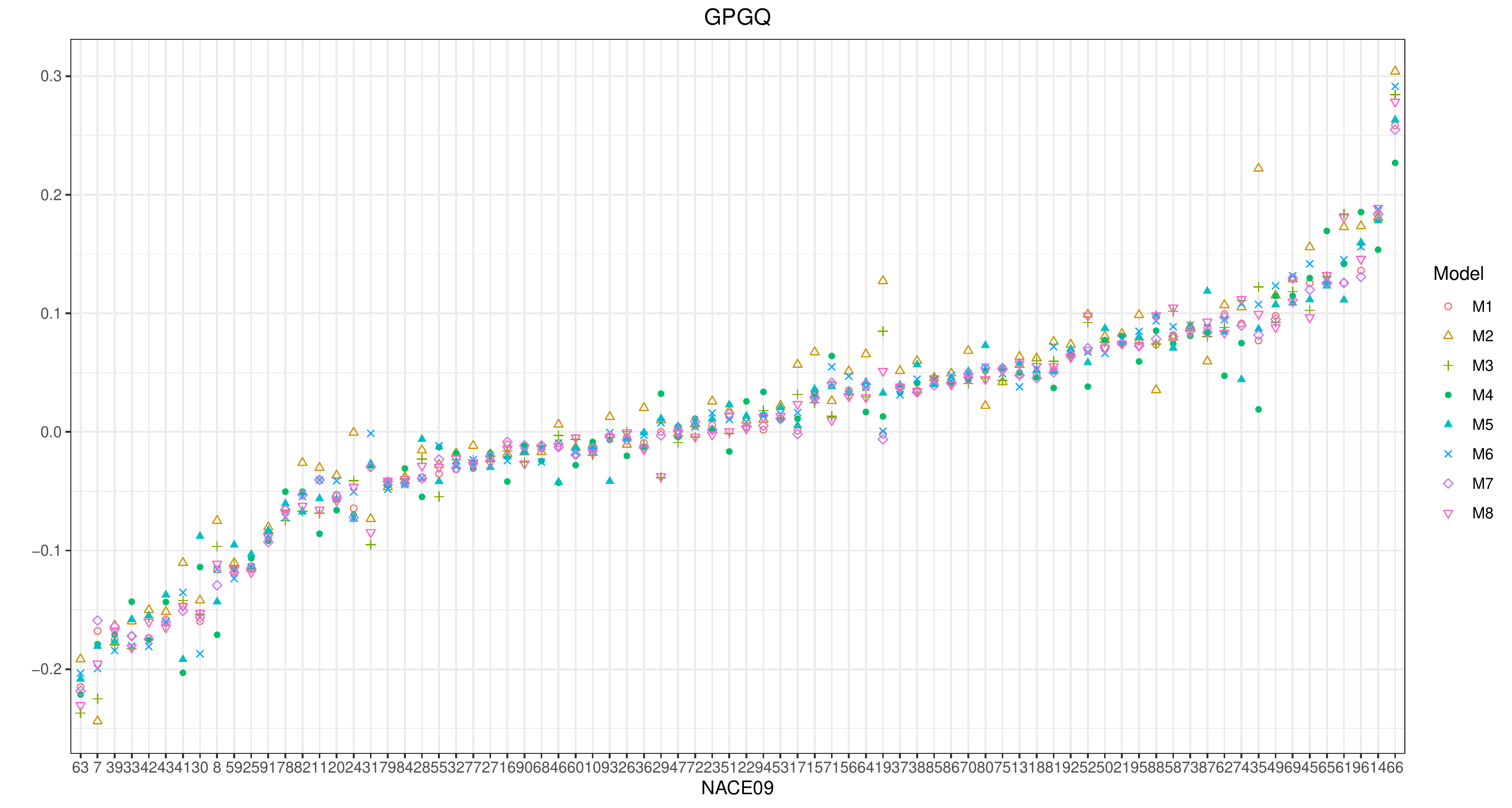}
\caption {\label{GPGq} $\widehat{GPG}_{Qd}$ for all the models considered and for each activity $d$ defined by the NACE09.}
\end{center}
\end{figure}

\begin{figure}[!ht]
\begin{center}
\includegraphics[width=180mm,height=90 mm]{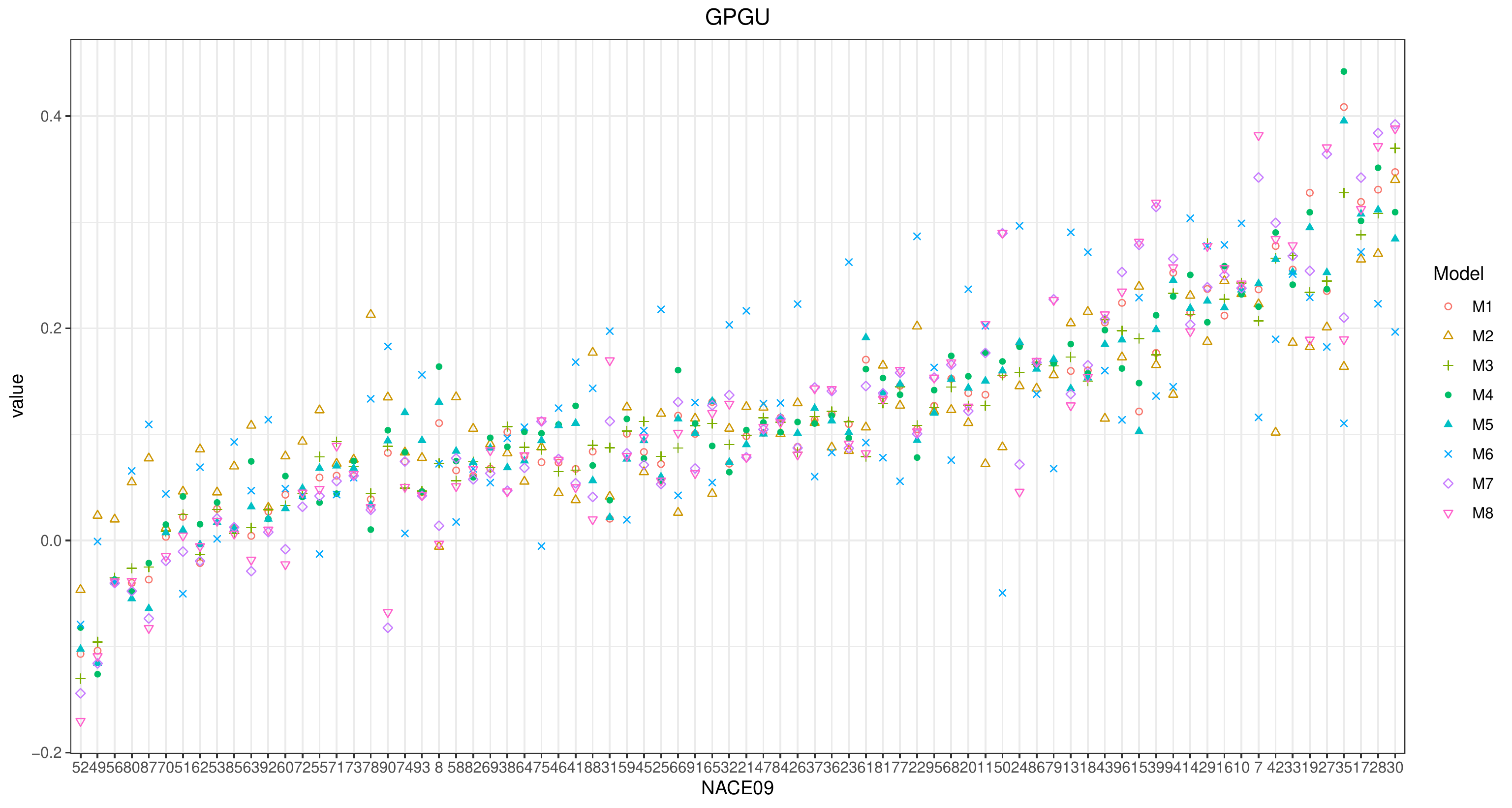}
\caption {\label{GPGu} $\widehat{GPG}_{Ud}$ for all the models considered and for each activity $d$ defined by the NACE09.}
\end{center}
\end{figure}

In order to explore differences between models, the estimated bias for each activity has been calculated for each candidate model and is shown in Figure \ref{bias}. The activities are sorted in increasing order of bias according to the $M1$ model. Figure \ref{bias} shows  important bias in some activities: $88=$ {S\textit{ocial work activities without accommodation}}, $90=$  { \textit{Creative, arts and entertainment activities}} or $39=$ {\textit{Remediation activities and other waste management services}}, among others. 
Moreover, the bias corrected estimator of the explained and unexplained parts of the $GPG$ for each economic activity and model, as shown in the expression (\ref{GPGestimado_d}), are shown in Figures \ref{GPGq}  and \ref{GPGu}, respectively,  sorting the activities by the values of $\widehat{GPG}_{Qd}$  and $\widehat{GPG}_{Ud}$ for $M1$.  These Figures show that while estimates for  $\widehat{GPG}_{Qd}$ are similar between models,  there are important differences in $\widehat{GPG}_{Ud}$ across models. 
%This fact points to the requirement to do model selection.
%to estimate the explained part of the pay gap, it is necessary to take into account the complexity of the model. We can see in Figure \ref{boxplot_bias} that a good alternative to M7 and M8 are M1 and M3, with less parameters to estimate. 

 The model selected, using the proposal described in Section 3, is $M1$. The parameter estimates for $M1$ are shown in Table \ref{coef}, which explains that the wage of a worker increase with  higher education, working in an enterprise with a national, UE or international market, and belonging to a large enterprise. The coefficients related to occupations show negative signs as {\it Professionals and managers} is the reference category. The results are consistent with those from other studies (Aláez et al. (2003), De la Rica (2008) and Moral-Arce et al. (2011)). \\

\begin{table}[h!]
\centering
\begin{tabular}{|lrrr|}

  \hline
 & Estimate & Std. Error & p-value \\ 
   \hline
   \hline
  Fixed effects&&&\\
  \hline
(Intercept)  & 2.299 & 0.033 & 0.000 \\ 
\hline
  Experience ($X_1$) & 0.007 & 0.001 & 0.000 \\
   \hline
   Age ($X_2$) & 0.005 & 0.000 & 0.000 \\ 
    \hline
Secondary ($X_3$)& 0.015 & 0.011 & 0.188 \\ 
  Higher ($X_4$) & 0.122 & 0.014 & 0.000 \\ 
   \hline
  Technic ($X_5$) & -0.248 & 0.015 & 0.000 \\
  Operators ($X_6$) & -0.439 & 0.016 & 0.000 \\ 
  Services and sales workers ($X_7$)& -0.504 & 0.019 & 0.000 \\ 
   Non-skilled workers ($X_8$)& -0.527 & 0.016 & 0.000 \\  
    \hline
  Short term contract ($X_9$) & -0.098 & 0.010 & 0.000 \\ 
     \hline
  Part time work  ($X_{10}$) & -0.017 & 0.013 & 0.210 \\ 
    \hline
  Between 20 to 49 workers ($X_{11}$) & -0.029 & 0.012 & 0.028 \\
  Between  50 to 99 workers ($X_{12}$)& 0.047 & 0.014 & 0.003 \\
    Between 100 to 199 workers ($X_{13}$) & 0.042 & 0.014 & 0.011 \\ 
  Between 200 to 499 workers ($X_{14}$)& 0.081 & 0.014 & 0.000 \\  
  More than 500 workers ($X_{15}$)& 0.113 & 0.016 & 0.000 \\ 
    \hline
  National Market ($X_{17}$)& 0.052 & 0.010 & 0.000 \\ 
  UE Market ($X_{18}$) & 0.076 & 0.018 & 0.001 \\
 International Market ($X_{19}$)& 0.072 & 0.016 & 0.000 \\ 
 \hline
 Random effects&&&\\
 \hline
 $\sigma_u^2$&0.021&&\\
 \hline

\end{tabular}
\caption{\label{coef} Coefficient estimates for men and for $M3$.}
\end{table}

Pay gap estimators have been obtained using the model $M1$ and  the proposal in Section 3. In Galicia, the  pay gap between men and women ($\widehat{ GPG}$) is estimated to be of 11.3\%, where  2.8\% is due to the observed quantify effect ($\widehat{GPG_{Q}}$) and 8.5\% to discrimination ($\widehat{ GPG_{U}}$), as Table \ref{tabla1} shows.  \\

\begin{table}[h!]
\centering
\begin{tabular}{|l|cc|}
  \hline
 &Estimation& CI 95\%\\ 
  \hline
  
  $\widehat{ GPG_Q}$& 0.028& (-0.018,0.07)\\
  $\widehat{ GPG_U}$ &0.085& (0.057,0.14) \\
     \hline
  $\widehat{ GPG}$ & 0.113&-\\
   \hline
\end{tabular}
\caption{\label{tabla1} Decomposition of gender pay gap in Galicia.}
\end{table}

\begin{figure}[!ht]
\begin{center}
\includegraphics[width=160mm,height=90 mm]{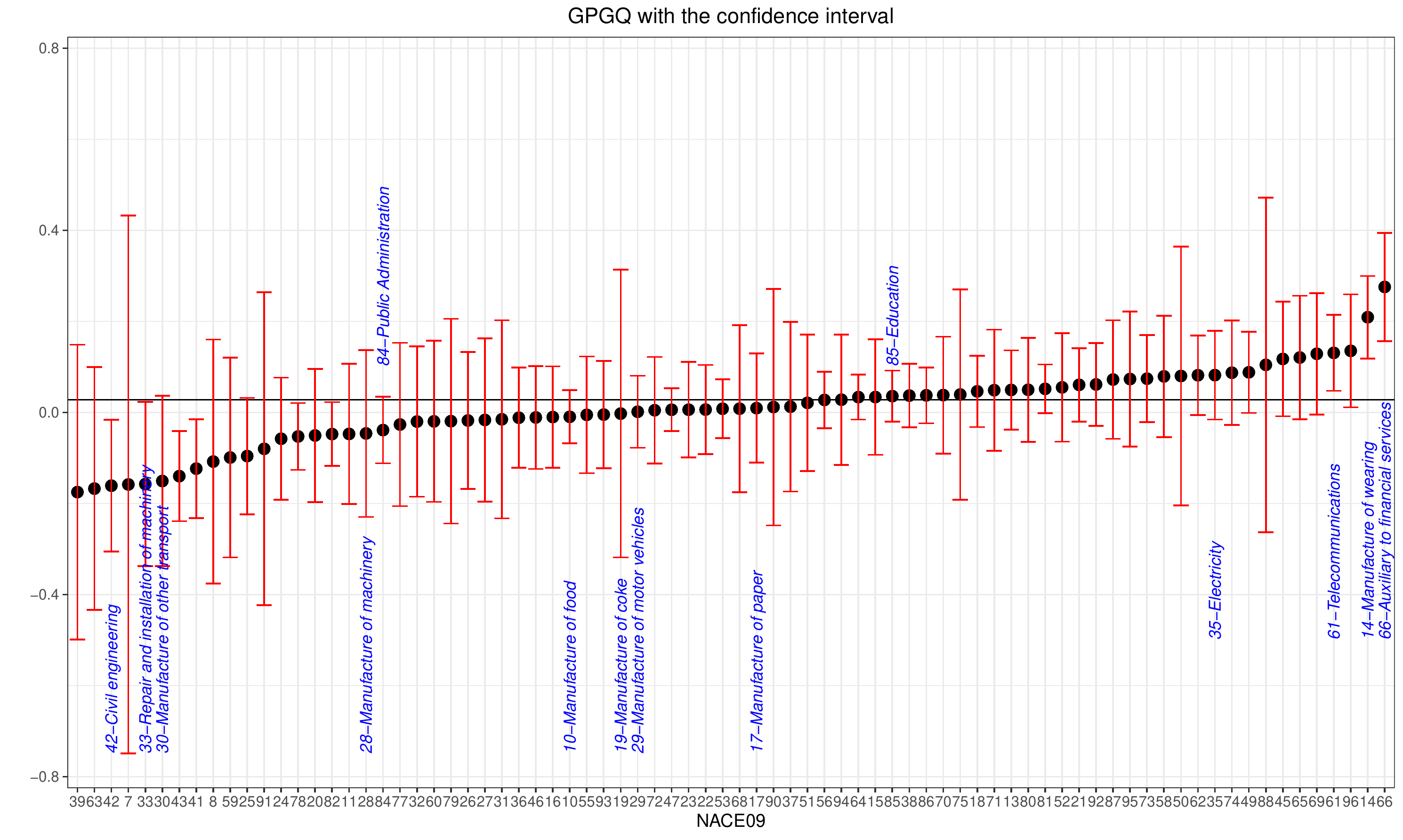}
\caption {\label{GPGq_confidence} $\widehat{GPG}_{Qd}$ with their confidence interval (95\%) and for each activity $d$ defined by the NACE09. The horizontal continuous line is the $\widehat{GPG}_{Q}$ for Galicia.}
\end{center}
\end{figure}

\begin{figure}[!ht]
\begin{center}
\includegraphics[width=160mm,height=90 mm]{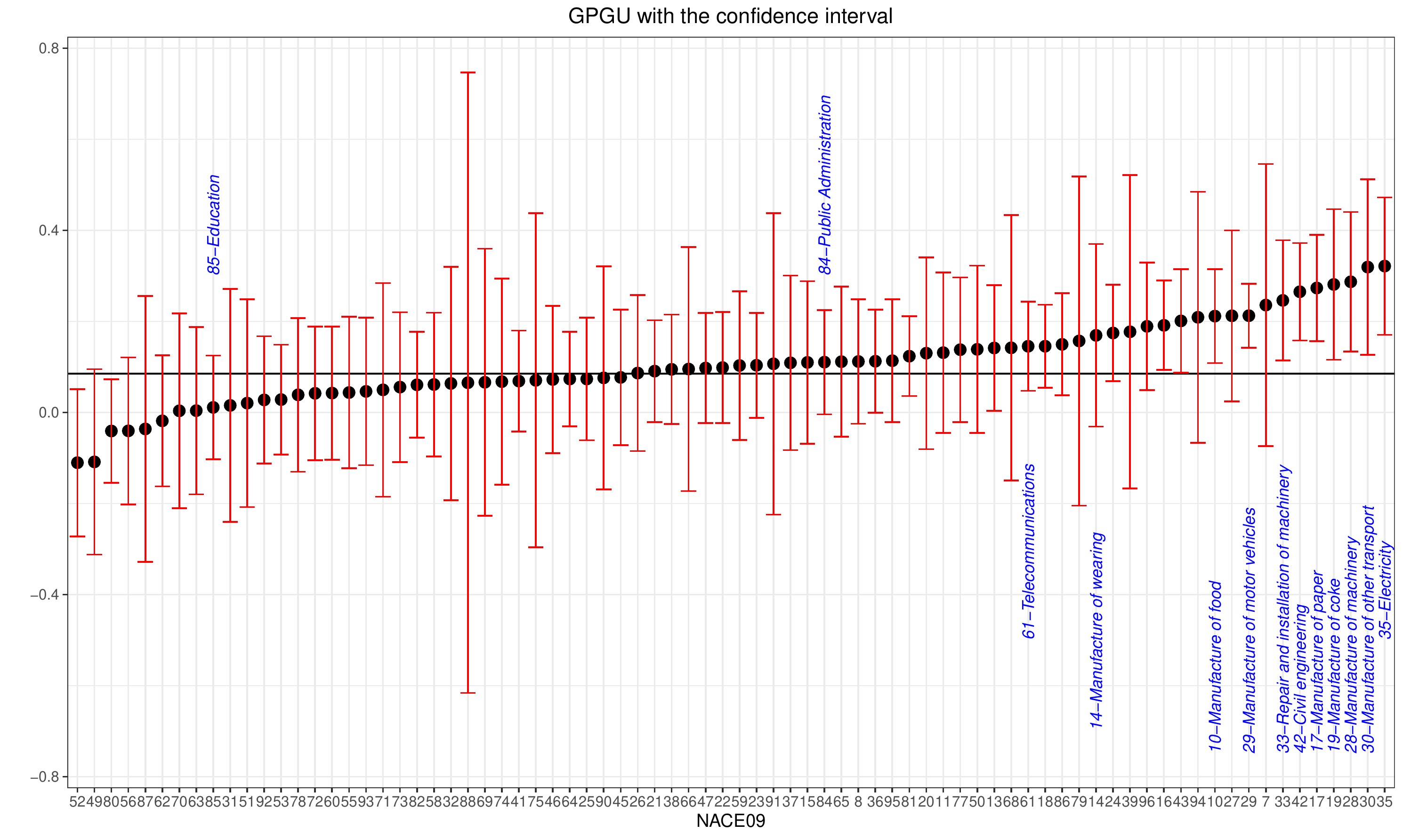}
\caption {\label{GPGu_confidence} Discrimination ($\widehat{GPG}_{Ud}$) with their confidence interval (95\%) and for each activity $d$ defined by the NACE09. The horizontal continuous line is the $\widehat{GPG}_{U}$ for Galicia.}
\end{center}
\end{figure}

Moreover, Figures \ref{GPGq_confidence} and \ref{GPGu_confidence} show  $\widehat{GPG}_{Qd}$ and $\widehat{GPG}_{Ud}$, respectively,  for each activity, $d$, with their confidence intervals. Activities are ordered by the increasing value of the  estimator.

% and allow us to determine if the problem of each activity derives from the differences in the characteristics of men and women or from the discrimination. With this figures, it is clearly observed that the size and nature of the problem of wage differences by sex varies according to activity. 
It is worth commenting on the results from activities 84 and 85,  which correspond to the {Public Administration} and {Education}. They are useful for validating the procedure  because no discrimination is expected in these cases as the Administration control the wages (Aláez et al., 2000, Moral-Arce et al., 2011). Figure \ref{GPGq_confidence} shows no significant differences between the characteristics of men and women, for activities 84 and 85; while Figure \ref{GPGu_confidence} shows, for 84 and 85, no significative differences from those of Galicia. The results are consistent with what would be expected.

On the other hand, the results show very high discrimination in eight activities: 30, 28, 19, 17, 42 , 33,  29 and 10, almost all of them are from the industrial sector (see Figure \ref{GPGu_confidence}); while differences between the characteristics due to gender are not significantly higher for these activities than for Galicia (Figure \ref{GPGq_confidence}). 


Taking as reference the $GPG$ of the region of Galicia, Table \ref{tabla} shows the activities with higher $\widehat{GPG_d}$ than Galicia, sorted from highest to lowest. The activity at the top is \textit{Electricity, gas, steam and air conditioning supply} and it is also one of the activities where there are more discrimination.
%The value of the discrimination is significative higher than the mean of Galicia that is 0.085, and is not in its confidence interval [0.17,0.47]. Also, 
The second activity at the top is \textit{Manufacture of wearing apparel}, and this is an example that high wage differences are not necessarily accompanied by discrimination. This activity is very important in the Galician business sector and employs approximately 8,600 workers, $1\%$ of the total number of employed people in Galicia. % In this case the difference between the characteristics due to gender is higher than that of Galicia.
On the other hand, note that only ten activities out of  74 have positive discrimination significantly higher than the region's mean. These activities employ $8.2\%$ of the workers in the region. It is also interesting to note that there are two activities in which discrimination is significantly negative (in favor of women). These are \textit{ Warehousing and support activities for transportation} and \textit{ Security and investigation activities} that belong to the service sector where women have a more significant presence than men.
\\
% Continuing with the description of Table \ref{tabla}, activity  61 is, also, an example that high wage differences is explained by the different characteristics of men and women. \\
%It can be appreciated that the activities enjoy the most desirable situation, because both differences and low wage discrimination coincide with respect to the average of Galicia. In the opposite extreme are activities 66 and 96 that share the most complicated situation with levels above the average.

% There are $34$ activities that have a higher pay gap , but in $8$ of this activities the pay gap it is not significatively different of 0. Then we obtain 26 economic activities with the highest gap, 18 belong to the industrial sector, 2 to the construction sector and the remaining 6 to the services sector. This agrees with the results of other studies where the industrial sector appears as the sector in which there is greater discrimination.

%Finally, in order to validate the goodness of the decompose of the pay gap, a simulation experiment that imitate the real case is included in the next section.

\newpage

\section{Simulations} 
 
The aim of this section is  to validate the novel proposal and compare it with the classical approach of OB. Moreover, the estimation error that results when important explanatory variables are omitted, is also illustrated. \\
  
The simulated experiment imitates SES real data. The 30 activities which have the largest sample sizes  and the  explanatory variables in Table \ref{variables} are considered. The sample sizes take values from 21 to 111 for men  and 7 to 82 for women.
The model used to generate the data is defined as follows:

\begin{eqnarray}\label{modelsimul}
\begin{array}{lll}
Y_{gdi} &=& X_{1gdi} \beta_{1gd}+X_{3gd}\beta_{2gdi}+X_{4dg}\beta_{3gd}+X_{5gdi}\beta_{4gd}+X_{6gdi}\beta_{5gd}+X_{7gdi}\beta_{6gd}+X_{8gdi}\beta_{7gd}+\epsilon_{gdi} \\
 && g\in \{m, w\}; \quad\ d=1,...,30; \quad\ i=1, \ldots, n_{gd};
 \end{array}\end{eqnarray}

\noindent where  $\epsilon_{gdi} \sim N(0,\sigma_{gd}^2)$, with 
$\sigma_{gd}^2=0.1$
and $\beta_{jgd}$, $j=1, \ldots,7$  are the estimated values using model $M6$ . \\

Note that model (\ref{modelsimul}) is much more complex than any of the models in the candidate list. The generating model includes a fixed parameter for each activity and explicative variable, giving a model with 8*30=240 parameters; being  the candidate models much more simple.

For each $d$, $GPG_d$, $GPG_{Qd}$ and $GPG_{Ud}$ are obtained using (1), (2) and (5) and are assumed to be true values. 

%These values are to be estimated from the working models shown in Table \ref{models2} and the algorithm proposed in Section \ref{Models}, taking $I=100$ iterations.\\

The models in  Table \ref{models2}  combine fixed and random effects, as in the real case. Estimators for GPG components and  each model are obtained using proposal in Section 3.
%  $u_d$ is the random effect associated to the economic activity; $u_{dj}, j=1, 2,$ is the random effect associated to the interaction economic activity and education;  $u_{dk}$ $k=1, 2, 3, 4,$ is the random effect associated to economic activity and occupation; and $v_d$ is the random slope associated to the economic activity and the variable $X_1$. We obtain, also, the estimators for the model of Oaxaca-Blinder and for the model selected by the $xGAIC$.

{\tiny \begin{table}[h!]
\centering
\begin{tabular}{|llccc|}
\hline
 Label   &		Fixed effects	 &		Random Effects &Fixed parameters (nº)&Random parameters (nº)	\\
\hline
$MS1$&$X_1$,$X_3$-$X_{8}$&$u_d$ &7&1\\
$MS2$&$X_1$,$X_3$-$X_{8}$&$v_d$&7&1\\
$MS3$&$X_1$,$X_3$-$X_{8}$&$u_d$, $v_d$&7&2\\
$MS4$&$X_1$,$X_3$-$X_{8}$&$u_{dj}$&7&2\\
$MS5$&$X_1$,$X_3$-$X_{8}$&$u_{dk}$&7&4\\
$MS6$&$X_1$,$X_3$-$X_{8}$, $X_{19}$-$X_{27}$&-&16&0	\\
$MS7$&$X_1$,$X_3$-$X_{8}$, $X_{28}$-$X_{56}$ &-&36&0\\
$MS8$&$X_1$,$X_3$-$X_{8}$, $X_{28}$-$X_{56}$&$v_d$&36&1\\

\hline
\end{tabular}
\caption{\label{models2} Models used to fit the data.}
\end{table}

}

%to measure the error of these estimators, we calculate the

Now,  the empirical mean square error (EMSE) of $\widehat{GPG}_{Qd}$ and $\widehat{ GPG}_{Ud}$ for all the models and for each activity $d$ are as follows:
$$
 EMSE(\widehat{GPG}_{Qd})=\frac{1}{100} \sum_{i=1}^{100}\left(\widehat{GPG}_{Qd}^{(i)}-{GPG}_{Qd}\right)^2
$$
$$
 EMSE(\widehat{GPG}_{Ud})=\frac{1}{100} \sum_{i=1}^{100}\left(\widehat{GPG}_{Ud}^{(i)}-{GPG}_{Ud}\right)^2 
$$
where $\widehat{GPG}_{Qd}^{(i)}$  and $\widehat{GPG}_{Ud}^{(i)}$ are the estimations for each iteration $i=1,\ldots, 100$.\\

Finally, we calculate the confidence intervals for $ GPG_{Qd}$ and ${GPG_{Ud}}$ from (\ref{ICQU}) and give the percentage of times that the true values $ GPG_{Qd}$ and ${GPG_{Ud}}$ are in the corresponding confidence interval.

Tables \ref{tabla7}  and \ref{tabla8} present the results of the simulations, EMSE values and coverage rates of confidence intervals, respectively. Table \ref{tabla7} shows the mean of the EMSE of $\widehat{ GPG}_{Qd}$ and $\widehat{ GPG}_{Ud}$ obtained from the models $MS1$ to $MS8$;  the OB  model and the model selected by xGAIC (XG) are also included.
The lefthand part of the table corresponds to the  model with all the explanatory variables and the righthand side shows the results when the information relative to education ($X_3$ and $X_4$) is removed. 
%The best model, in terms of EMSE, is $MS5$.  
%In the real case the best model is $M1$, one of the reasons of the difference between the real case and the simulations maybe that in the simulations we only use $30$ economic activities. 
%Now, if you study the robustness of the estimators in the absence of relevant information (see the right part of Table \ref{tabla7}), it can be seen that the EMSE increases in all the models, but especially in the OB and MS6 model.\\

The best models, in terms of EMSE, are $MS5$ and $XG$. When relevant information is missed out the error increases in all the models, but the increase is higher for OB and the MS6 model.
The conclusions are similar regarding the coverage rates of 95\% confidence intervals in Table \ref{tabla8}.
% Again, the models proposed have a better behavior than the model of Oaxaca-Blinder.  It is also interesting to verify that if we eliminate important variables from the estimation of the models, such as $X_3$ and $X_4$, the influence is much smaller in the models proposed than the model of Oaxaca-Blinder, because the confidence intervals have greater coverage, although less than $95\%$.
From these results, it can be said that the proposed methodology is robust to the potential effects of differences in unobserved variables; to be precise, much more than the OB approach.\\

 In addition, in Tables \ref{tabla7} and  \ref{tabla8},  you can see that the models selected by the $xGAIC$ have a good behavior, because they are among those that have a lower EMSE and a greater coverage. When the model is fitted with all the variables,  the $xGAIC$ selects the MS5 model $82\%$ of times, $15\%$ of times $MS4$ and $3\%$ $MS8$.  When $X_3$ and $X_4$ are simultaneously removed from the working model, the $xGAIC$ selects the model $MS5$ $98\%$ of times and $2\%$ the model $MS8$.
 \\
Finally, this study highlights the gain of our methodology with respect to the OB methodology, especially for estimating ${ GPG}_{Ud}$.

\begin{table}[ht]
\centering
\begin{tabular}{|c|cc|cc|}
        \hline
        && & Without $X_3$, $X_4$&\\

  \hline
 & $GPG_Q$ & $GPG_U$  & $GPG_Q$ & $GPG_U$ \\ 
  \hline
 OB&  0.0049 & 0.0067 &0.0060 &0.0089\\ 
 MS1 & 0.0046 &   0.0045 &0.0054 &  0.0057\\ 
  MS2 & 0.0057 &  0.0066 & 0.0067 &   0.0076  \\ 
   MS3 & 0.0060 &  0.0048 &0.0066 &   0.0058   \\ 
   MS4 & 0.0045 &   0.0044 & 0.0054 &   0.0057  \\ 
 MS5 & 0.0044 &   0.0044 & 0.0052 &   0.0054  \\ 
  MS6 & 0.0049 &   0.0067 &0.0061 & 0.0079 \\ 
  MS7 & 0.0046 &   0.0048 &  0.0053 &  0.0060 \\ 
   MS8 & 0.0060 &  0.0049 & 0.0065 & 0.0060 \\ 
   XG&0.0045&0.0043&0.0052&0.0054\\
  \hline
\end{tabular}\caption{\label{tabla7} EMSE of $GPG_Q$ and $GPG_U$ for the model with and without  $X_3$ and $X_4$. OB is the Oaxaca-Blinder model and XG is the model selected by the $xGAIC$}
\end{table}

\begin{table}[ht]
\centering
\begin{tabular}{|c|cc|cc|}
        \hline
        && &Without $X_3$, $X_4$&\\
  \hline
 & $GPG_Q$ & $GPG_U$  & $GPG_Q$ & $GPG_U$ \\ 
  \hline
  OB &  86.4 & 55.0&80.1 & 43.7 \\ 
  MS1 & 86.4 & 78.9 & 80.5 & 73.7 \\ 
  MS2 & 83.6 & 57.4 & 79.0  & 54.6 \\ 
  MS3 & 88.7  & 81.4 & 84.2  & 77.8\\ 
  MS4 & 90.1 & 83.0 & 80.5  & 73.7\\ 
  MS5 & 90.8 & 84.5 & 85.1  & 81.5\\ 
  MS6 & 86.4 & 55.0& 80.2  & 45.9  \\ 
  MS7 & 86.4 & 81.7 & 80.2 & 76.2 \\ 
  MS8 & 88.7  & 83.2 & 83.9  & 79.4 \\ 
  XG& 90.7&84.4&85.0&81.0  \\

   \hline
\end{tabular}\caption{\label{tabla8} Coverage of the $GPG_Q$ and $GPG_U$ confidence intervals for the model with and without  $X_3$ and $X_4$. OX is the Oaxaca-Blinder model and XG is the model selected by the $xGAIC$}
\end{table}

%\begin{figure}[!ht]
%\begin{center}
%\includegraphics[width=180mm,height=90 mm]{MSE_withoutX2}
%\caption {\label{MSE_simul} MSE for  $\widehat{ GPG}_{Qd}^{B}$ and  $\widehat{ GPG}_{Ud}^{B}$ for the different models in the simulations and when we eliminate $X2$.}
%\end{center}
%\end{figure}

% Wed Sep  5 19:52:03 2018
\begin{table}[ht]
\centering
\begin{tabular}{|r|r|rrr|rrr|}
  \hline
NACE09 & $\widehat {GPG}_{d}$ &  $\widehat{ GPG}_{Qd}$ & $\widehat{GPG}_{Qd}^{-}$ & $\widehat{GPG}_{Qd}^{+}$ &  $\widehat {GPG}_{Ud}$& $\widehat{GPG}_{Ud}^{-}$& $\widehat{GPG}_{Ud}^{+}$ \\ 
  \hline
35 & 0.40 & 0.08 & -0.02 & 0.18 & 0.32 & 0.17 & 0.47 \\ 
 14 & 0.35 & 0.21 & 0.12 & 0.30 & 0.17 & -0.03 & 0.37 \\ 
 96 & 0.32 & 0.14 & 0.01 & 0.26 & 0.19 & 0.05 & 0.33 \\ 
19 & 0.28 & -0.00 & -0.32 & 0.31 & 0.28 & 0.12 & 0.45 \\ 
  17 & 0.27 & 0.01 & -0.11 & 0.13 & 0.27 & 0.16 & 0.39 \\ 
 61 & 0.27 & 0.13 & 0.05 & 0.21 & 0.15 & 0.05 & 0.24 \\ 
  28 & 0.25 & -0.05 & -0.23 & 0.14 & 0.29 & 0.13 & 0.44 \\ 
  65 & 0.24 & 0.12 & -0.02 & 0.26 & 0.11 & -0.05 & 0.28 \\ 
 94 & 0.22 & 0.03 & -0.12 & 0.17 & 0.21 & -0.07 & 0.48 \\ 
 50 & 0.21 & 0.08 & -0.20 & 0.36 & 0.14 & -0.05 & 0.32 \\ 
 29 & 0.21 & 0.00 & -0.08 & 0.08 & 0.21 & 0.14 & 0.28 \\ 
 45 & 0.20 & 0.12 & -0.01 & 0.24 & 0.08 & -0.07 & 0.23 \\ 
 10 & 0.20 & -0.01 & -0.07 & 0.05 & 0.21 & 0.11 & 0.31 \\ 
 13 & 0.19 & 0.05 & -0.04 & 0.14 & 0.14 & 0.00 & 0.28 \\ 
 18 & 0.19 & 0.05 & -0.03 & 0.12 & 0.15 & 0.05 & 0.24 \\ 
  69 & 0.19 & 0.13 & -0.00 & 0.26 & 0.07 & -0.23 & 0.36 \\ 
86 & 0.19 & 0.04 & -0.02 & 0.10 & 0.15 & 0.04 & 0.26 \\ 
  27 & 0.19 & -0.02 & -0.20 & 0.16 & 0.21 & 0.02 & 0.40 \\ 
  95 & 0.19 & 0.07 & -0.07 & 0.22 & 0.11 & -0.02 & 0.25 \\ 
  16 & 0.18 & -0.01 & -0.12 & 0.10 & 0.19 & 0.09 & 0.29 \\ 
  81 & 0.18 & 0.05 & -0.00 & 0.11 & 0.12 & 0.04 & 0.21 \\ 
 21 & 0.17 & 0.06 & -0.02 & 0.14 & 0.09 & -0.02 & 0.20 \\ 
  74 & 0.16 & 0.09 & -0.03 & 0.20 & 0.07 & -0.16 & 0.29 \\ 
 30 & 0.16 & -0.15 & -0.34 & 0.04 & 0.32 & 0.13 & 0.51 \\ 
 88 & 0.15 & 0.10 & -0.26 & 0.47 & 0.07 & -0.62 & 0.75 \\ 
 15 & 0.14 & 0.03 & -0.09 & 0.16 & 0.11 & -0.07 & 0.29 \\ 
  73 & 0.14 & 0.07 & -0.02 & 0.17 & 0.06 & -0.11 & 0.22 \\ 
  37 & 0.14 & 0.01 & -0.17 & 0.20 & 0.11 & -0.08 & 0.30 \\ 
  58 & 0.14 & 0.08 & -0.05 & 0.21 & 0.06 & -0.10 & 0.22 \\ 
 68 & 0.13 & 0.01 & -0.18 & 0.19 & 0.14 & -0.15 & 0.43 \\ 
  38 & 0.13 & 0.04 & -0.03 & 0.11 & 0.09 & -0.03 & 0.22 \\ 
  75 & 0.12 & 0.04 & -0.19 & 0.27 & 0.07 & -0.30 & 0.44 \\ 
  79 & 0.12 & -0.02 & -0.24 & 0.21 & 0.16 & -0.20 & 0.52 \\ 
% 64 & 0.11 & 0.03 & -0.02 & 0.08 & 0.07 & -0.03 & 0.18 \\ 
 %77 & 0.11 & -0.03 & -0.21 & 0.15 & 0.14 & -0.02 & 0.30 \\ 
 %24 & 0.11 & -0.06 & -0.19 & 0.08 & 0.17 & 0.07 & 0.28 \\ 
 %23 & 0.11 & 0.01 & -0.10 & 0.11 & 0.10 & -0.01 & 0.22 \\ 
 %36 & 0.10 & -0.01 & -0.12 & 0.10 & 0.11 & -0.00 & 0.23 \\ 
 %22 & 0.10 & 0.01 & -0.09 & 0.10 & 0.10 & -0.02 & 0.22 \\ 
%47 & 0.10 & 0.01 & -0.04 & 0.05 & 0.10 & -0.02 & 0.22 \\ 
 %71 & 0.10 & 0.05 & -0.08 & 0.18 & 0.05 & -0.18 & 0.28 \\ 
       \hline
       \end{tabular}
\caption{\label{tabla} Observed gap, $\widehat{ GPG}_{Qd}^{B}$ and $\widehat{ GPG}_{Ud}^{B}$ with the confidence interval for the 33 activities with the highest observed gaps.}
\end{table}

\newpage

\section{Conclusions }

The first contribution of this paper, and the most important from the theoretical point of view, is the derivation of a novel approach to estimate the components of the gender pay gap in small areas. This methodology is consistent with that used by Eurostat (Leythienne and Ronkowski, 2018), it also generates confidence intervals for the explained and unexplained part of the GPG, includes a bias correction and considers linear mixed models and a selection of models option. Moreover, this novel approach is robust against potential differences in unobserved variables.

The second, and most important contribution from the practical point of view, is that the application of the methodology to estimate wage discrimination in economic activities in Galicia reveals important differences in the region and among activities. 

 In Galicia, $25\%$ of the pay gap is explained by the differences between characteristics and $75\%$ is due to discrimination.  In other words, if there were no discrimination and the characteristics of men and women had been paid at the same prices, the wage differential would be reduced to 2.8\%.
 
Regarding the activities, a high pay gap due to differences in the characteristics is shown in such activities as 66 and 14, while a high pay gap due to discrimination is shown in other activities, such as 28, 19, 17, among others.
%proposed estimates the salary differences by sex in the different economic activities and check if there was discrimination or not, with confidence intervals that allow us to see the differences from the mean of Galicia. With this, we can see that there are important differences between the activities. 

The existence of salary differences by sex continues to be a real problem, according to the results obtained. The $GPG$ indicator, together with the explained and unexplained gap by economic activities, allow for a better identification and interpretation of the causes of the gender pay gap. As a consequence, policy actions can be better targeted.
% These results are different by activities, where salary differences are very important in some activities, like  28-Manufacture of machinery and equipment  with a discrimination of 29\%. 

Finally,  to determine whether two workers have the same qualification and value, useful information, such as the total work experience, knowledge of languages or the awards or publications obtained, among many others, is still missing. The inclusion of this information as an explanatory variable would likely derived in more accurate pay gap estimators.

 \section{Appendix}
 
 Tables \ref{nace091} and \ref{nace092} show the codes and the description of the Statistical Classification of Economic Activities in the European Community (NACE09 Rev.2) at the division level, excluding the primary sector.
 
\begin{table}[ht]
\centering
\begin{tabular}{|l|l|}
  \hline
NACE09 & Description \\ 
  \hline

6	&	Extraction of crude petroleum and natural gas	\\
7	&	Mining of metal ores	\\
8	&	Other mining and quarrying	\\
9	&	Mining support service activities	\\
10	&	Manufacture of food products	\\
11	&	Manufacture of beverages	\\
12	&	Manufacture of tobacco products	\\
13	&	Manufacture of textiles	\\
14	&	Manufacture of wearing apparel	\\
15	&	Manufacture of leather and related products	\\
16	&	Manufacture of wood and of products of wood and cork, except furniture	\\
17	&	Manufacture of paper and paper products	\\
18	&	Printing and reproduction of recorded media	\\
19	&	Manufacture of coke and refined petroleum products	\\
20	&	Manufacture of chemicals and chemical products	\\
21	&	Manufacture of basic pharmaceutical products and pharmaceutical preparations	\\
22	&	Manufacture of rubber and plastic products	\\
23	&	Manufacture of other non-metallic mineral products	\\
24	&	Manufacture of basic metals	\\
25	&	Manufacture of fabricated metal products, except machinery and equipment	\\
26	&	Manufacture of computer, electronic and optical products	\\
27	&	Manufacture of electrical equipment	\\
28	&	Manufacture of machinery and equipment n.e.c.	\\
29	&	Manufacture of motor vehicles, trailers and semi-trailers	\\
30	&	Manufacture of other transport equipment	\\
31	&	Manufacture of furniture	\\
32	&	Other manufacturing	\\
33	&	Repair and installation of machinery and equipment	\\
35	&	Electricity, gas, steam and air conditioning supply	\\
36	&	Water collection, treatment and supply	\\
37	&	Sewerage	\\
38	&	Waste collection, treatment and disposal activities; materials recovery	\\
39	&	Remediation activities and other waste management services	\\
41	&	Construction of buildings	\\
42	&	Civil engineering	\\
43	&	Specialised construction activities	\\
45	&	Wholesale and retail trade and repair of motor vehicles and motorcycles	\\
46	&	Wholesale trade, except of motor vehicles and motorcycles	\\
47	&	Retail trade, except of motor vehicles and motorcycles	\\
49	&	Land transport and transport via pipelines	\\
50	&	Water transport	\\
51	&	Air transport	\\
52	&	Warehousing and support activities for transportation	\\
53	&	Postal and courier activities	\\
55	&	Accommodation	\\
56	&	Food and beverage service activities	\\
\hline

       \hline
       \end{tabular}
\caption{\label{nace091} List of codes for the NACE09 at division level, excluding the primary sector.}
\end{table}

\begin{table}[ht]
\centering
\begin{tabular}{|l|l|}
  \hline
NACE09 & Description \\ 
  \hline

58	&	Publishing activities	\\
59	&	Motion picture, video and television programme production, sound recording and music \\
60	&	Programming and broadcasting activities	\\
61	&	Telecommunications	\\
62	&	Computer programming, consultancy and related activities	\\
63	&	Information service activities	\\
64	&	Financial service activities, except insurance and pension funding	\\
65	&	Insurance, reinsurance and pension funding, except compulsory social security	\\
66	&	Activities auxiliary to financial services and insurance activities	\\
68	&	Real estate activities	\\
69	&	Legal and accounting activities	\\
70	&	Activities of head offices; management consultancy activities	\\
71	&	Architectural and engineering activities; technical testing and analysis	\\
72	&	Scientific research and development 	\\
73	&	Advertising and market research	\\
74	&	Other professional, scientific and technical activities	\\
75	&	Veterinary activities	\\
77	&	Rental and leasing activities	\\
78	&	Employment activities	\\
79	&	Travel agency, tour operator and other reservation service and related activities	\\
80	&	Security and investigation activities	\\
81	&	Services to buildings and landscape activities	\\
82	&	Office administrative, office support and other business support activities	\\
84	&	Public administration and defence; compulsory social security	\\
85	&	Education	\\
86	&	Human health activities	\\
87	&	Residential care activities	\\
88	&	Social work activities without accommodation	\\
90	&	Creative, arts and entertainment activities	\\
91	&	Libraries, archives, museums and other cultural activities	\\
92	&	Gambling and betting activities	\\
93	&	Sports activities and amusement and recreation activities	\\
94	&	Activities of membership organisations	\\
95	&	Repair of computers and personal and household goods	\\
96	&	Other personal service activities	\\
97	&	Activities of households as employers of domestic personnel	\\
98	&	Undifferentiated goods- and services-producing activities of private households for own use\\
99	&	Activities of extraterritorial organisations and bodies	\\
       \hline
       \end{tabular}
\caption{\label{nace092} List of codes for the NACE09 at division level, excluding the primary sector (continuation).}
\end{table}

\section{Bibliography}

\begin{description}

\item

Akaike, H. (1973). Information theory and the maximum likelihood principle. In \textit{International Symposium on Information Theory}, pages 267-281, Budapest. Akademiai Kiado.
\item[]

Aláez, R., Ullibarri, M. (2000). Discriminación salarial por sexo: un análisis del sector privado y sus diferencia regionales en España. ICE, \textit{Revista de Economía}, 789, 117-138.
\item[]

Aláez, R., Longás, J.C., Ullibarri, M. (2001). Visualising gender wage differences in the European Union. \textit{Gender, Work and Organization}.
\item[]

Aláez, R., Longás, J.C., Ullibarri, M. (2003). Diferencias salariales en España: un análisis sectorial/regional. \textit{Investigaciones regionales}, 3, 5-24.
\item[]

Anastasiade, M.C. and Tillé, Y. (2017a). Decomposition of gender wage inequalities through calibration: Application to the Swiss structure of earnings survey. \textit{Survey Metodology}, 43:211-234.
\item[]

Anastasiade, M.C. and Tillé, Y. (2017b). Gender wage inequalities in Switzerland: the public versus the private sector. \textit{Statistical Methods and Applications}, 26 (2) 293-316. 
\item[]

Battese, G. E., Harter, R. M. and Fuller, W. A.  (1988). An error-components model for prediction of county crop areas using survey and satellite data. \textit{Journal of the American Statistical Association}, 83(401),28-36.
\item[]

Berk, R., Brown, L. Buja, a. Zhang, K. and Zaho L. (2013). Valid post-selection inference.  \textit{The Annals of Statistics}, 41(2), 802-837.\\
\item[]

Blinder, A.S. (1973). Wage Discrimination: Reduced Form and Structural Estimates. \textit{Journal of Human Resources}, 8(4), 436-455.
\item[]

De la Rica, S., Dolado, J.J., Llorens, V. (2008). Ceilings or floors: gender wage gaps by education in Spain. \textit{J. Popul Econ},  21, 751-778
\item[]

Dempster,A. P., Rubin D. B. and Tsutakawa R.K. (1981). Estimation in covariance components models. \textit{Journal of the American Statistical Association}, 76, 341-353.
\item[]

Fernández, M., Montuenga, V. M., Romeu, A E. (2000). Wage differencials and non competitive behaviour in the Spanish industry labor market. \textit{Documentos de traballo análise económica}, n 9, IDEGA. 
\item[]

%Fisher, R.A. (1992). On the mathematical of theoretical statistics. \textit{Phil Trans Roy Soc Lond, Series A }, 222, 209-268.
%\item[]

%Cotton J (1988). On the Decomposition of Wage Differentials. \textit{Review of Economics and Statistics}, 70(2), 236-243.
%\item[]
Fan, Y., and Li, R. (2012).  Variable selection in linear mixed effects models. \textit{Annals of statistics}, 40(4), 2043-2068.
\item[]

Fortin, N., Lemieux, T., and Firpo, S. (2011). Decomposition methods in economics. In Ashenfelter, O. and Card, D., editors, \textit{Handbook of Labor Economics}, volume 4, 1-102. Elsevier.
\item[]

Graf, M. , Mar\'{\i}n, M. and Molina, I. (2019). A generalized mixed model for skewed distributions applied to small area estimation. \textit{TEST}, 28, 565-597.
\item[]

Hlavac, Marek (2018). oaxaca: Blinder-Oaxaca Decomposition in R. R package version 0.1.4. https://CRAN.R-project.org/package=oaxaca.
\item[]

Jann B (2008). The Blinder-Oaxaca Decomposition for Linear Regression Models. \textit{Stata Journal}, 8(4), 453-479.
\item[]

Jiang, J. (2007). {\it Linear and Generalized linear Mixed Models and Their Applications}. Springer Series in Statistics.
\item[]

Juhn, C., Murphy, K. M., and Pierce, B. (1993). Wage inequality and the rise in returns to skill. \textit{Journal of Political Economy}, 101(3):410-442.
\item[]

Lee, J. D., Sun, D. L., Sun, Y. and Taylor J. E. (2016). Exact post-selection inference with application to the lasso. \textit{The Annals of Statistics}, 44(3), 907?927
\item[]

Leythienne, D. abd Ronkowski P. (2018). A decomposition of the unadjusted gender pay gap using Structure of Earnings Survey data. {\it Population and social conditions. Statistical Working Papers}. Publications Office of the European Union.
\item[]

Lombardía, M.J., López-Vizcaíno, E., and Rueda, C. (2017). Mixed generalized akaike information (xGAIC) for small area models. \textit{Journal of Royal Statistical Society Series A}, 180, 1229-1252.
\item[]

Moral-Arce, I., Sperlich, S., Fernandez-Sáenz, A.I., Roca, M.J. (2011). Trends in the Gender Pay Gap in Spain: A Semiparametric Analysis. \textit{J. Labor Res}, 33, 173?195.
\item[]

%Neumark D (1988). Employers Discriminatory Behavior and the Estimation of Wage Dis- crimination. \textit{Journal of Human Resources}, 23(3), 27-295.
%\item[]

%Neuman, S. and Oaxaca,R. (2004). Wage decompositions withe selectivity-corrected wage equations: A methodological note. \textit{Journal of Economic Inequality}, 2, 3-10.
%\item[]

Oaxaca R.L. (1973). Male-Female Wage Differentials in Urban Labor Markets.\textit{International Economic Review}, 14(3), 693-709.
\item[]

Oaxaca R.L., Ransom M.R. (1994). On the discrimination and the decomposition of wage differentials. \textit{Journal of Econometrics}, 61, 5-21.
\item[]

Oaxaca R.L., Ransom M.R. (1998). Calculation of approximate variances for wage decomposition differentials. \textit{Journal of Economic and Social Measurement}, 81(1), 154-157.
\item[]

Opsomer, J., Claeskens, G., Ranalli, M., Kauermann, G., and Breidt, F. (2008). Non-parametric small area estimation using penalized spline regression. {\it Journal of Royal Statistical Society
Series B}, 70, 265-286.
\item[]

%Oaxaca R.L., Ransom M.R. (1999). Identification in Detailed Wage Decompositions. \textit{Review of Economics and Statistics}, 81(1), 154-157.
%\item[]

%Reimers C.W. (1983). Labor Market Discrimination Against Hispanic and Black Men. \textit{Review of Economics and Statistics}, 65(4), 570-579.
%\item[]
Popli, G.K. (2013). Gender wage differentials in Mexico: a distributional approach. \textit{Journal of the Royal Statistical Society Series A}, 176(2), 295-319.
\item[]

Rao, J.N.K. and Molina, I. (2015). {\it Small area estimation}, Second Edition. J. Wiley, Hoboken (NJ).
\item[]

Rueda, C. and Lombard\'{\i}a, M.J. (2012). Small area semiparametric additive monotone models. {\it Statistical Modelling}, 12, 527-549.
\item[]

Tibshirani, R.J., Taylor J. Lockhart R. and Tibshirani R. (2016). Exact Post-Selection Inference for Sequential Regression Procedures. {\it Journal of the American Statistical Association},  111(514), 600-620.\\
\item[]

%Weichselbaumer, D., and Winter-Ebmer, R. (2005). A meta-analysis of the international gender wage gap.
%\textit{Journal of Economic Surveys}, 19(3), 479-511.

\item[]

\end{description}

\end{document}